\documentclass[fleqn,usenatbib]{mnras}

\usepackage{newtxtext,newtxmath}

\usepackage[T1]{fontenc}
\usepackage[]{ulem}
\usepackage[]{xcolor}
\usepackage[]{pdflscape}
\usepackage{tablefootnote}

\DeclareRobustCommand{\VAN}[3]{#2}
\let\VANthebibliography\thebibliography
\def\thebibliography{\DeclareRobustCommand{\VAN}[3]{##3}\VANthebibliography}
\usepackage{siunitx}
\DeclareSIUnit \parsec {pc}
\DeclareSIUnit \astrounit {au}

\def\apj{ApJ}
\def\apjl{ApJL}
\def\aap{A\& A}
\def\nat{Nature}

\def\mnras{MNRAS}

\def\apjs{ApJS}

\def\mnras{{MNRAS}}

\def\alamenos#1{$^{-#1}$}

\newcommand{\effturb}{\epsilon_{\rm ff-turb}}

\newcommand{\Machffturb}{{\mathcal M}_{\rm ff-turb}}

\newcommand{\Msun}{$M_\odot$}    %
\usepackage{graphicx}	
\usepackage{amsmath}	

\newcommand{\lo}    {$L_{\odot}$}
\newcommand{\mo}    {$M_{\odot}$}

\newcommand{\LJeans}{L_{\mathrm{J}}}
\newcommand{\LJog}{L_{\mathrm{Jog}}}

\newcommand{\LamJeans}{\lambda_{\mathrm{J}}}
\newcommand{\LamJog}{\lambda_{\mathrm{Jog}}}
\newcommand{\LamJeansfluc}{\lambda_{\mathrm{J_p}}}
\newcommand{\LamJeanscero}{\lambda_{\mathrm{J}_0}}

\newcommand{\Menv}{M_{\mathrm{env}}}
\newcommand{\Mfluct}{M_{\mathrm{\delta_r}}}
\newcommand{\MJeans}{M_{\mathrm{J}}}
\newcommand{\MJog}{M_{\mathrm{Jog}}}

\newcommand{\rhoeff}{\rho_{\mathrm{eff}}}
\newcommand{\rhoefflarson}{\rho_{\mathrm{eff, Larson}}}
\newcommand{\rholarson}{\rho_{\mathrm{Larson}}}
\newcommand{\rhofluct}{\rho_{\mathrm{fluct}}}
\newcommand{\sigmaturb}{\sigma_{\mathrm{ff-turb}}}
\newcommand{\supa}  {$^\mathrm{a}$}
\newcommand{\supb}  {$^\mathrm{b}$}
\newcommand{\supc}  {$^\mathrm{c}$}
\newcommand{\supd}  {$^\mathrm{d}$}
\newcommand{\supe}  {$^\mathrm{e}$}
\newcommand{\supf}  {$^\mathrm{f}$}
\newcommand{\supg}  {$^\mathrm{g}$}

\title[Tidal forces in star-forming regions]{The effect of tidal forces on the Jeans instability criterion in star-forming regions}

\author[R. Zavala-Molina et al.]{
Rafael Zavala-Molina$^{1}$\thanks{E-mail: r.zavala@irya.unam.mx},  Javier Ballesteros-Paredes$^{1}$, Adriana Gazol$^1$, and Aina Palau$^1$
\\
$^{1}$Instituto de Radioastronom\'{i}a y Astrof\'{i}sica, Universidad Nacional Aut\'{o}noma de M\'{e}xico, PO Box 3-72, 58090 Morelia, Michoac\'{a}n, Mexico\\
}

\date{Accepted XXX. Received YYY; in original form ZZZ}

\pubyear{2022}

\begin{document}
\label{firstpage}
\pagerange{\pageref{firstpage}--\pageref{lastpage}}
\maketitle

\begin{abstract}
Recent works have proposed the idea of a tidal screening scenario, in which gravitationally unstable fragments in the vicinity of a protostar will compete for the gas reservoir in a star-forming clump. In this contribution, we propose to properly include the action of an external gravitational potential in the Jeans linear instability analysis as proposed by Jog. We found that an external gravitational potential can reduce the critical mass required for the perturbation to collapse if the tidal forces are compressive or increase if are disruptive. Our treatment provides (a) new mass and length collapse conditions; (b) a simple equation for observers to check whether their observed fragments can collapse;  and (c) a simple equation to compute whether collapse-induced turbulence can produce the levels of observed fragmentation. Our results suggest that, given envelopes with similar mass and density, the flatter ones should produce more stars than the steeper ones. If the density profile is a power-law, the corresponding power-law index separating these two regimes should be about 1.5. We finally applied our formalism to 160 fragments identified within 18 massive star-forming cores of previous works. We found that considering tides, 49\% of the sample may be gravitationally unstable and that it is unlikely that turbulence acting at the moment of collapse has produced the fragmentation of these cores. Instead, these fragments should have formed earlier when the parent core was substantially flatter.
\end{abstract}




\section{Introduction}

The Jeans length/mass instability criterion \citep{Jeans02} has been widely used to determine whether a parcel of gas is stable against  collapse. It is obtained by making a perturbation analysis of the Poisson equation along with the equations of the hydrodynamics. When this analysis is performed, one of the main assumptions is that the gravitational potential due to the external mass distribution is negligible. Thus, only the gravitational potential due to the perturbation is considered. However, there might be physical configurations where tidal forces may play a role. \\

Whether tidal forces over molecular clouds coming from the whole galaxy, neighboring molecular clouds, or even if they are relevant is still a matter of debate. For instance, \citet{Ballesteros-Paredes+09a, Ballesteros-Paredes+09b} have shown that tidal forces due to an effective external potential can compress or disrupt molecular clouds and even affect the star formation efficiency. In addition, \citet{Meidt+18} found that the galactic potential may be relevant in feeding molecular cloud turbulence, a result that agrees with \citet{Liu+21}. The last authors, for instance, argue that the self-gravity of molecular clouds in the S0-type galaxy NGC 4429 may be as relevant as the external pull from their parent galaxy.\\

In contrast to the results previously mentioned, some authors have found that the galactic potential does not contribute to the tidal energy of molecular clouds. In order to show this, \citet{Suarez+12} used toy models while \citet{Ramirez-Galeano+22} analyzed numerical simulations of the Cloud Factory suite \citep{Smith+20}. It should be pointed out that although these works found that galactic tides are not relevant, it is not necessarily the case of tides due to neighbor molecular clouds. For example, \citet{Mao+20} and \citet{Ramirez-Galeano+22}  found that the boundedness of a molecular cloud may depend on the molecular clouds mass distribution around it. In addition, \citet{Ramirez-Galeano+22} found that tidal stresses from neighboring molecular cloud complexes, rather than the galactic potential (as mentioned in the previous paragraph), may feed interstellar turbulence. However, to complicate the situation,  \cite{Ganguly+22} found that tidal stresses from nearby clouds are utterly irrelevant in molecular clouds. \\ 

In order to understand this variety of results in apparent contradiction, it is important to understand the different configurations analyzed in each of these works. For example, although 
a galactic potential may not contribute to the energy budget of a molecular cloud \citep{Suarez+12, Ramirez-Galeano+22}, it may be relevant to HI clouds and, thus, to how molecular clouds are piled up \citep{Ramirez-Galeano+22}. As a result, tidal stresses between molecular clouds may be relevant when spiral arms are present \citep{Mao+20, Ramirez-Galeano+22}, and may be negligible far away from the arms, explaining the results of \citet{Ganguly+22}. Indeed, the latter authors based their results on simulations from the SILCC-Zoom suite \citep{Seifried+18}. These simulations have a vertical gravitational potential of the stellar disk. However, they do not contain a full galactic potential (dark matter halo, stellar bulge, galactic stellar disk, and stellar spiral arms). In summary, tides working on molecular clouds may depend upon the position of the cloud within their host galaxy \citep[as reported by ][]{Meidt+18, Liu+21}. In addition, it could be the reason why the \citet{Larson81} relations depend upon the position of the cloud within their host galaxy \citep{Colombo+14}. \\

On the sub-parsec scale, it has been found that simple molecular cloud potentials at scales of a few pc are not relevant in compressing or tearing apart their $l\lesssim 0.1$~pc molecular cloud cores \citep{Ballesteros-Paredes+09a}. Nonetheless, the results by \citet{Mao+20} and \citet{Ramirez-Galeano+22} suggest that more complex molecular cloud gravitational potentials could have a tidal effect on their inner cores, a situation yet to be explored more carefully.   \\

On even smaller scales, tidal forces have been invoked to understand whether they can prompt or inhibit the formation of protostars within dense molecular cloud cores. In this sense, \citet[][from now on, LH18]{Lee_Hennebelle18b}  have formulated a tidal screening scenario as a physical mechanism that explains why the initial mass function of the stars (IMF) has a peak around 0.2~\Msun. In this scenario, tidal forces due to a pre-existing protostar and its associated envelope determine whether a density fluctuation around the protostar can collapse and the radius inside which gravitational fragmentation is prevented, setting up the available mass to be accreted by the central object \citep[see also][]{Lee+20}. A similar analysis, but considering only {the tidal tensor}, was performed by \citet[][from now on, CT20]{Colman_Teyssier20}. In this analysis, the collapse conditions for density fluctuations and the possibility of achieving them in realistic protostellar environments are crucial. \\

In the present work, we explore a different approach to collapse conditions than those utilized in LH18 and CT20 and examine its potential implications. We base our analysis on the criterion delineated by \citet{Jog13}, who re-derived the Jeans criterion by considering an external gravitational potential's contribution to the momentum equation. 
This modification shows that an external potential can reduce the Jeans mass needed for a cloud to gravitationally collapse if the external potential is compressive or increase it if the potential is disruptive.
In section \ref{sec:tools} we expose the physical problem to tackle and the theoretical tools needed.  In section \ref{sec:stab_jog_env} we study the stability of density fluctuations using the modified criterion and we also introduce new collapse conditions. In section \ref{sec:applications} we apply our theoretical development to observations of fragmenting dense cores. And finally, in section \ref{sec:discussion} we provide a discussion of our results and the conclusions.  



\section{Physical configuration and theoretical tools}\label{sec:tools}

\subsection{The tidal tensor of the envelope and the Larson core} \label{sec:lee_and_colman}

In order to investigate the physical conditions allowing for the  gravitational collapse of  density fluctuations embedded in the envelope of a protostellar object, we consider a spherical envelope whose density varies radially following a power-law  profile with the form
%
\begin{equation}\label{eq:rhoenv}
      \rho_{\rm env}(r) = \rho_0 \left( \frac{r}{r_0}\right)^{-\alpha}.
\end{equation}

It is well known that the gravitational collapse of an unstable sphere of gas produces an object with an isothermal density profile, $\rho \propto r^{-2}$ \citep{Larson69}. However, it could be possible that density fluctuations are formed earlier, when the density field is still flat, being the reason why we leave $\alpha$ undetermined. 
The mass contained by the envelope inside a sphere of radius $r$ is then 
 %
 %
\begin{align} \label{eq:masaenv}
\begin{split}
    M_{\rm env} (r) &= \int_0^{r} \rho_{\rm env}(r') 4 \pi r'^2 dr' \\
    &= 4\pi \rho_0 r_0^{\alpha} \frac{r^{3-\alpha}}{3-\alpha} = \frac{4 \pi}{3-\alpha} \rho_{\rm env}(r) r^3.
\end{split}
\end{align}
{The previous equation admits profiles with $\alpha<3$ however, note that} finite energy solutions require $\alpha < 2.5$. Following LH18 and CT20, we characterize the density fluctuations by $\eta$, a parameter that determines the amplitude of the fluctuation  
\begin{equation}
    \rho_{\rm fluct} (r) = \eta \rho_{\rm env}(r)  = \eta \rho_0  \left( \frac{r}{r_0}\right)^{-\alpha}.
\label{eq:den_pert_potencias}
\end{equation}
%
In contrast with LH18, we consider any density fluctuation, regardless of whether its mass is larger than the mass of Larson's core. This situation allows the formation of stars at earlier times as well as the formation of brown dwarfs.\\

%
%
The force balance at the perturbation position determines the stability of the density fluctuation. The force includes the contribution of the total gravitational field experienced by the fluctuation and the contribution of the thermal pressure gradient. The gravitational field within the volume of the density fluctuation is given by the potential produced by the central protostar's mass, the mass of the spherical envelope, also centreed at $r=0$, and the mass of the density fluctuation itself. A valuable tool to account for the external contributions to this gravitational field is the tidal tensor, which
is defined as\footnote{Note that the sign in the last equation is changed with respect to other works (e.g. CT20), where another convention is used.}
\begin{equation}
  \tau_{ij} = - \frac{\partial^2\Phi}{\partial x_i x_j},
  \label{eq:tensor}
\end{equation}
where $\Phi$ is the total gravitational potential and $x_i, x_j=(1,2,3)$ are the components of any orthogonal
coordinate system. 
%
In a spherical coordinate system and assuming radial symmetry, $\tau$ becomes diagonal

\begin{align*}
\tau = \begin{pmatrix}
-\frac{\partial^2 \Phi}{\partial r^2} & 0 & 0 \\
0 & -\frac{1}{r}\frac{\partial \Phi}{\partial r} & 0 \\
0 & 0 & -\frac{1}{r}\frac{\partial \Phi}{\partial r}
\end{pmatrix}
= \begin{pmatrix}
-\frac{\partial F_r}{\partial r} & 0 & 0 \\
0 & -\frac{F_r}{r} & 0 \\
0 & 0 & -\frac{F_r}{r}
\end{pmatrix}. \\
\end{align*}

The contribution to the tidal tensor due to a central protostar with mass $M$ is, thus, 

%
\begin{align} 
    \tau^{\rm c} 
    = \frac{GM}{r^3} \
    \begin{pmatrix}
    2 &  0 &  0 \\
    0 & -1 &  0 \\
    0 &  0 & -1 
    \end{pmatrix},
    \label{eq:marea_centro}
\end{align}
while the tidal tensor of an envelope with a power-law density profile as described by equation (\ref{eq:rhoenv}) is\footnote{The detailed derivation of eq.~(\ref{eq:marea_env}) can be found in \cite{Masi07}, where the change in nature of the radial tidal force, from tensile to compressive,  is explored. The author points out that, inside a spherical mass distribution (not necessarily homogeneous, nor a power-law profile), the tidal force at a given position $r_0$ becomes compressive if $\rho(r_0) > 2 \rho_{\rm mean}/3$, 
%
%
where 
\begin{equation}
\rho_{\rm mean} = \frac{\int_{0}^{r_0} 4\pi \rho_{\rm env}(r) r^2 dr} {4\pi r_0^3/3} \nonumber
\end{equation}
is the mean density of the material enclosed by a sphere with radius $r_0$, \citep[see eq.~(20) and the discussion that follows][]{Masi07} .}

\begin{align} 
\begin{split}
    \tau^{\rm env} &= \frac{4 \pi G}{3-\alpha}\rho_0 \left( \frac{r}{r_0}\right)^{-\alpha} \ \begin{pmatrix} 
    \alpha -1 &  0 &  0\\
            0 & -1 &  0\\
            0 &  0 & -1 
    \end{pmatrix}\nonumber\\
    &= \frac{4 \pi G}{3-\alpha}\rho_{\rm env} \ \begin{pmatrix} 
    \alpha -1 &  0 &  0\\
            0 & -1 &  0\\
            0 &  0 & -1 
    \end{pmatrix}.
    \end{split}  \\
    \label{eq:marea_env}
\end{align}

As commented out by different authors \citep[e.g.,][]{Masi07, Colman_Teyssier20}, when the concavity of the potential is negative,
\begin{equation}
    \frac{\partial^2 \Phi}{\partial x^2} < 0,
\end{equation}
the tidal force is disruptive. From eq.~(\ref{eq:marea_env}), we notice that the two tangential components of the tensor are always compressive. Thus, the interest is focused on whether the radial component is compressive or disruptive. For the density distribution considered in the present work, the transition from a tensile to a compressive radial tidal force {regime} occurs at $\alpha =1$. This implies that for relatively flat profiles, i.e., relatively early in the collapse process, Jeans stable density fluctuations could be unstable due to the tidal effect of the envelope. Conversely, tidal forces stabilize otherwise unstable density fluctuations when the envelope has a relatively sharp density distribution. These possibilities naturally raise the question of how gravitational instability's critical length (mass) is modified when the tidal forces are considered. This problem has been addressed by \cite{Jog13}, {whose} analysis is outlined in the next section.  

\subsection{Jog length and mass}\label{sec:jog}

%
%
%
%
%
%
By performing a linear perturbation analysis of a 
{portion} of gas within an external gravitational potential, \citet{Jog13} could infer an equivalent to the Jeans length and mass, $\LJeans,$ $\MJeans$. The new quantities, which we call Jog length and mass, $\LJog$, $\MJog$, are related to the former by
%
\begin{equation}
        \lambda_{\rm Jog} = \lambda_{\rm J} \frac{1}{(1 - \rho_{\rm eff}/ \rho)^{1/2}},
         \label{eq:jog_lenght}
\end{equation}
and
\begin{equation}
    M_{\rm Jog}= M_{\rm J} \frac{1}{( 1 - \rho_{\rm eff}/\rho)^{3/2}}, 
    \label{eq:jog_mass}
\end{equation}
where $\rho$ is the density of the portion of fluid whose gravitational stability is being evaluated, 
\begin{equation}
    \rho_{\rm eff} = T_0/4\pi G
    \label{eq:rhoeff_T0}
\end{equation}
is an effective density\footnote{Note that the effective density $\rhoeff$, being just an  analogous to the external tidal force, can be negative or positive.}, and
%
\begin{equation}
    T_0 = \tau_{rr}= - \frac{\partial^2 \Phi}{\partial r^2} 
  \label{eq:T0}
\end{equation}
is the tidal force per unit distance in the $r$ direction. 
%
%
%
For $T_0 > 0$, the concavity of the gravitational potential is negative, and the tidal effect is disruptive. Furthermore, if $0 < \rhoeff < \rho$, the denominator in eqs.~(\ref{eq:jog_lenght}) and (\ref{eq:jog_mass}) becomes less than unity, and the critical length/mass for instability is larger relative to the Jeans value. In the limit $\rhoeff \to \rho$, 

\begin{equation}
    \frac{L_{\rm Jog}}{L_{\rm Jeans}} \rightarrow \infty \qquad \text{and} \qquad \frac{M_{\rm Jog}}{M_{\rm Jeans}} \rightarrow \infty,  
\end{equation}
and there is no solution for the Jog's length/mass. In other words, disruptive tidal forces sufficiently strong  ---compared with the size of the density fluctuation--- will never allow the collapse of the density fluctuation.  \\

On the other hand, if $\rhoeff<0$~ ($T_0 <0$), the concavity is positive, so the denominator in eq.~(\ref{eq:jog_lenght}) and (\ref{eq:jog_mass}) is larger than one, and therefore, the critical length/mass for instability decreases relative to the Jeans values. Now, at the limit $T_0 \rightarrow -\infty$
%
\begin{equation}
    \frac{L_{\rm Jog}}{L_{\rm Jeans}} \rightarrow 0 \qquad \text{and} \qquad \frac{M_{\rm Jog}}{M_{\rm Jeans}} \rightarrow 0.
\end{equation}
In other words, if the tidal forces are very compressive, the Jog's length/mass can become much less than their classical Jeans equivalents, making the collapse of density fluctuations much easier.\\

Note that the previous analysis, as the Jeans analysis, is one-dimensional, implying that in 3D, it is strictly applicable only for isotropic external fields, which is not the case for the configuration studied in this work. However, as discussed by \citet{Jog13}, the tidal fields in the other directions are always compressive, and their contributions are of the same order of magnitude as the radial one. Then, instability criteria found by using eqs.~(\ref{eq:jog_lenght}) and (\ref{eq:jog_mass}) will differ by a factor of a few compared to full 3D criteria. Our focus on the radial component effects is consistent with the one-dimensional character of the analysis. \\

Also, it is worth pointing out that Jog's analysis has some properties that makes it a powerful tool for treating the current problem. Since it arises from linearizing the mass, momentum, and Poisson equations for a portion of a fluid undergoing tidal forces, there are two important consequences. First, tides from different components, such as the spherical envelope or the Larson core, can be included in the tensor (\ref{eq:T0}). Second, there is no need to add additional terms to consider the thermal support or the density fluctuation's self-gravity since it already includes them. 
The present formalism allows us to avoid certain approximations made in previous approaches. More specifically, the pure tidal tensor analysis \citep[e.g.,][]{Renaud+11, Renaud+14, Colman_Teyssier20} does not include the potentially stabilizing effect of the thermal support nor the fluctuation self-gravity, which requires further approximations. Additionally, Jog's analysis does not require the standard approximation in theories that employ the density PDF in a turbulent gas \citep[e.g.][]{Padoan_Nordlund02, Krumholz_McKee05, Lee_Hennebelle18b}, which does not distinguish whether high-density gas is spatially connected.

%
\section{Stability of density fluctuations under tidal stresses}\label{sec:stab_jog_env}
In the present section, we use the formalism described in \S\ref{sec:tools} to determine the physical conditions leading to the gravitational collapse of density fluctuations in a spherical envelope with a power-law density profile. For simplicity, we will only include tides arising from the envelope without accounting for the Larson core since, in general terms, its contribution is substantially smaller, as we will argue in \S\ref{sec:larson}. 
\subsection{Jog mass of a density fluctuation within a power-law spherical envelope}\label{sec:jog_env}
%
%
%
By using eq.~(\ref{eq:rhoeff_T0}) and the first component of the tensor $\tau^{\rm env}$ in eq.~(\ref{eq:marea_env}), the effective density due to a power-law density distribution can be written as
\begin{align}
    \rho_{\rm eff} = \frac{\alpha-1}{3- \alpha} \rho_0 \left( \frac{r}{r_0}\right)^{-\alpha}.
    \label{eq:rho_eff_powerlaw}
\end{align}
For a fluctuation occurring at $r=r_{\rm p}$, the ratio between the effective density (given by eq.~[\ref{eq:rhoeff_T0}]) and the density of the fluctuation (eq.~[\ref{eq:den_pert_potencias}]), is then given by 
%
\begin{equation} \label{eq:cociente_densidad_equivalente}
    \frac{\rho_{\rm eff}}{\rho_{\rm fluct}} = \frac{\alpha-1}{\eta(3-\alpha)}.
\end{equation}
We have seen that if $\rho_{\rm eff}/\rho_{\rm fluct}<0$, then the tidal force is compressive. This occurs if  $\alpha < 1$. On the contrary, if $\alpha > 1$, then the tidal force is disruptive, and in this case, only fluctuations with 
%
\begin{equation}
    \eta > \frac{\alpha-1}{3-\alpha}
\end{equation}
can eventually be gravitationally unstable. Therefore, besides being sufficiently massive, density fluctuations in a power-law envelope should have a large enough density contrast to collapse.\\

Inserting now eq.~(\ref{eq:cociente_densidad_equivalente}) in eq.~(\ref{eq:jog_lenght}) and (\ref{eq:jog_mass}), the Jog length and mass become
\begin{equation} \label{eq:cociente_jog_jeans_1}
    \frac{\lambda_{\rm Jog}}{\lambda_{\rm J}} = \left[ 1- \left(\frac{\alpha-1}{3-\alpha}\right) \frac{1}{\eta}\right]^{-1/2}
\end{equation}
and
\begin{equation} \label{eq:cociente_jog_jeans}
 \frac{M_{\rm Jog}}{M_{\rm J}} = \left[ 1- \left(\frac{\alpha-1}{3-\alpha} \right) \frac{1}{\eta}  \right]^{-3/2}.   \\
\end{equation}

We notice again the three different regimes mentioned before.  (1) From eq.~(\ref{eq:cociente_densidad_equivalente}), when $\alpha = 1$ the ratio between the effective density and the density of the density fluctuation, $\rhoeff/\rhofluct\to$~0 goes to zero, and the Jog length/mass becomes equal to the Jeans length/mass. In other words, a sphere with a density profile  $\rho \propto r^{-1}$ does not exert tidal forces over density fluctuations embedded inside it, a situation well known in analysis of tidal forces \citep[e.g.,][]{Masi07}. (2) On the other hand, when $\alpha > 1$, the ratio $\rho_{\rm eff}/\rho_{\rm fluct}$ becomes larger than one, and the Jog's length and mass become larger than their Jeans equivalents. More specifically, steeper density profiles produce larger tides, making it harder for a density fluctuation to become gravitationally unstable. (3) Finally, when  $ \alpha < 1$, the ratio in eq.~(\ref{eq:cociente_densidad_equivalente}) becomes negative, and the Jog's length/mass decreases with respect to their classical Jeans counterparts, a configuration that will contribute to the gravitational collapse of the density fluctuation. \\

\begin{figure}
	\includegraphics[width=\columnwidth]{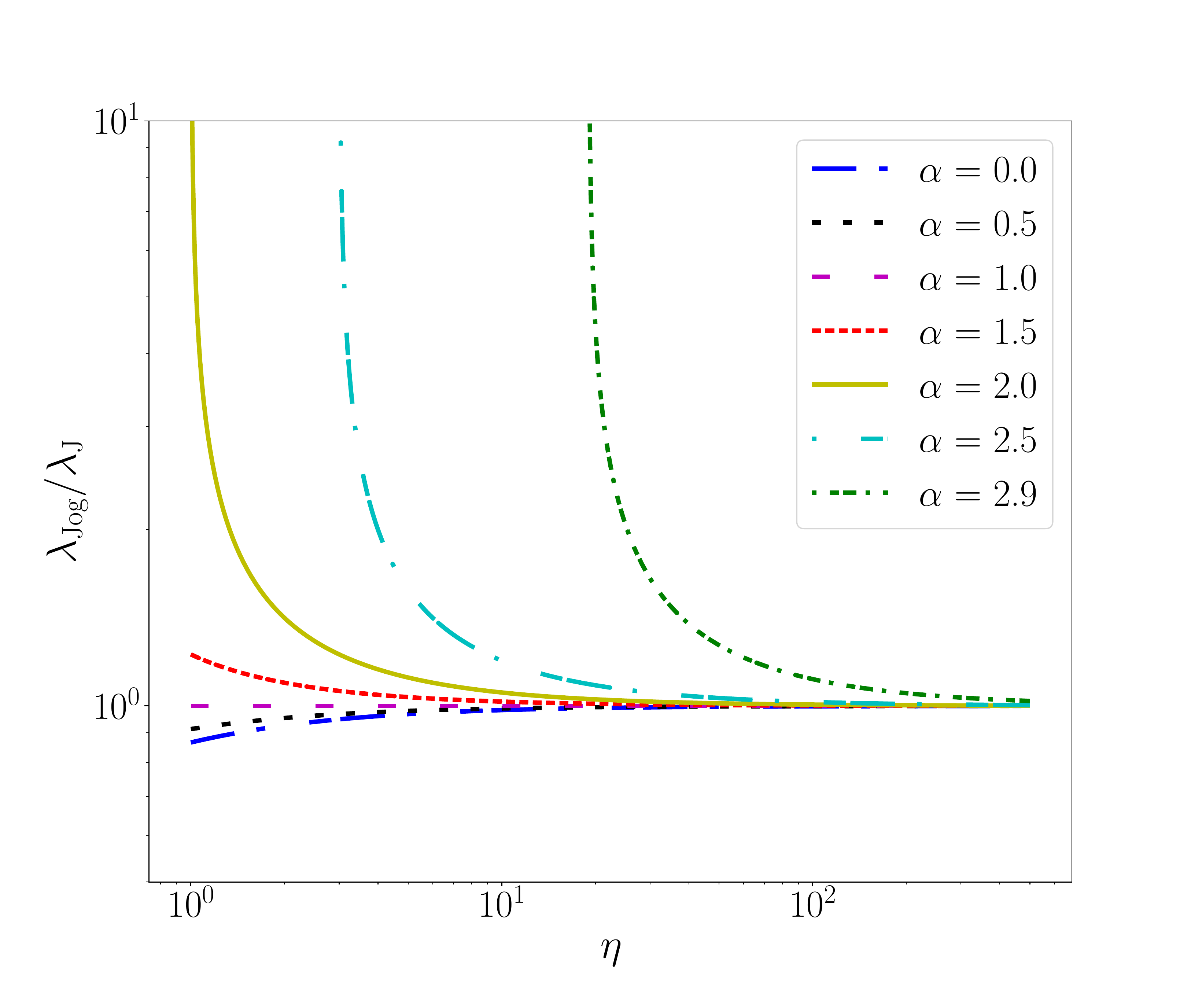} \\
	\includegraphics[width=\columnwidth]{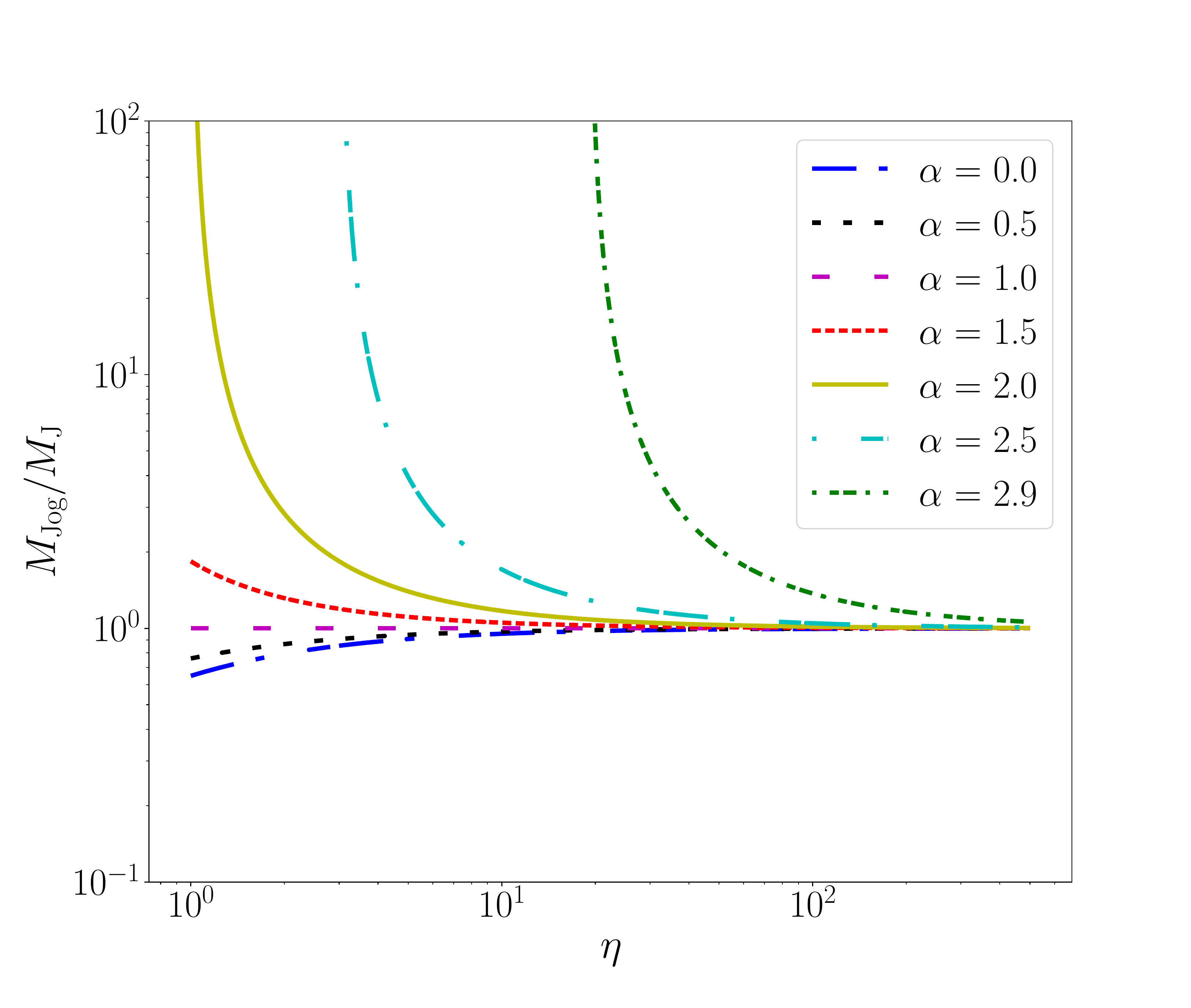}
    \caption{(a) Ratio $\lambda_{\rm Jog}/\lambda_{\rm J}$ (b) $\MJog/M_J$ as a function of $\eta$ for several power-law density profiles $\alpha$ of the envelope.}
    \label{fig:jog_length_mass}
\end{figure}

In Fig.~\ref{fig:jog_length_mass}, we show the ratio between Jog length (panel a) and Jog mass (panel b) over their corresponding Jeans length and mass, respectively, as a function of the density fluctuation's amplitude $\eta$. From this figure, we notice that the Jog and Jeans lengths (masses) are similar for large values of the density amplitude, $\eta$. This means that as the density amplitude $\eta$ becomes larger, the relevance of the self-gravity of the density fluctuation increases, reducing the relative relevance of tidal forces. In other words, as the density amplitude grows indefinitely, Jog's criterion approaches Jeans' criterion. 

It is important to stress that, as $\alpha$ becomes larger than $\sim$1.5, it becomes substantially harder for a density fluctuation to become self-gravitating.  This suggests, thus, that given two envelopes with similar mass and density, flatter massive envelopes should, in principle, be the progenitors of stellar clusters. In comparison, steeper envelopes should be responsible for the isolated mode of star formation\footnote{We thank the anonymous referee for pointing out this possibility}. 


As a second point, we also notice that the concavity of the ratios $\LamJog/\LamJeans$ and $\MJog/\MJeans$ changes from $\alpha < 1$ to $\alpha>1$. As we have just mentioned, a spherical envelope whose density changes as $\rho \propto r^{-1}$ does not exert tidal forces over clumps embedded inside it. Therefore, we get back the Jeans criterion. On the other hand, exponents strictly less than one indicate that Jog's length and mass are smaller than their Jeans counterparts. In other words, the envelope acts compressing, and density fluctuations whose mass does not reach their Jeans mass can collapse under our criterion. \\ 

Lastly, for density fluctuations immersed in density envelopes with $\alpha \geq 1$, Jog's length and mass are larger than their Jeans counterparts, making the gravitational collapse of the density fluctuation more unlikely to happen. In the case of an isothermal sphere, we note that Jeans and Jog's lengths differ mainly when the density contrast of the fluctuation is small. Jog's length and mass go to infinity when $\eta \rightarrow 1$, which makes us conclude that for density fluctuations embedded inside an isothermal sphere, it is extremely difficult to collapse when their density is slightly larger than that of the medium around them. This result can be interpreted as a scenario in which the variation of the envelope's density is so abrupt and the self-gravity of the density fluctuation so small that the envelope tears apart the density fluctuation. Therefore, we must again underline that the discrepancy between the Jeans criterion and our formulation is noticeable when the density jump of the density fluctuation is small. \\

Note that the behavior described above has been observed in numerical simulations of collapsing cores with different density profiles and turbulence levels \citep{Girichidis+12b}, where steeper profiles appear to be less prone to fragment. The physical interpretation of this fact in terms of differentiated tidal effects for distinct radial profiles has been outlined in \citep{Ntormousi_Hennebelle15}, where the shell-like stability of a homologously collapsing cloud is studied.  \\

Finally, it is important to recall that although spheres with steeper profiles than $\alpha = 2$ produce tidal forces that are increasingly important, reaching the limit $\alpha = 3$, where it is not possible to form a density fluctuation able to collapse,  profiles with $\alpha\ge 2.5$ are already unrealistic: at $\alpha=2.5$, gravitational energy already diverges, and for $\alpha=3$, the envelope mass is infinite. Thus, we can expect profiles not much larger than $\alpha\sim$~2 \citep[see also][]{Gomez+21}.

\subsection{New collapse conditions}
\label{sec:NewCollapseConditions}
From eqs.~(\ref{eq:cociente_jog_jeans_1}) and (\ref{eq:cociente_jog_jeans}), we could naively say that whenever $\alpha<3$, a large enough density fluctuation will always collapse because its Jeans and Jog's mass converge fast enough when the amplitude of the density fluctuation is large enough. Nevertheless, the important question is if the mass $\Mfluct$ of a density fluctuation with a density amplitude $\eta$ times the density of the power-law density sphere, can be as large as the Jog mass, $\MJog$, i.e., whether 
\begin{equation}\label{eq:collapse_cond}
   \frac{\Mfluct}{M_{\rm Jog}} \ge 1.
\end{equation}

Following the nomenclature in LH18, for a uniform spherical perturbation with size $\delta_r$ at a radius $r = r_{\rm p}$
\begin{equation}\label{eq:fluc_mass}
    \Mfluct= \frac{\pi}{6} \rhofluct \delta_r^3 = \frac{\pi}{6} \eta  \left( \frac{r_{\rm p}}{r_0}\right)^{-\alpha} \delta_r^3,
\end{equation}
while from eq. (\ref{eq:cociente_jog_jeans})
\begin{equation}
   M_{\rm Jog} = \frac{\pi}{6} \left[ 1- \left(\frac{\alpha-1}{3-\alpha} \right) \frac{1}{\eta} \right]^{-3/2} \rhofluct\ \LamJeansfluc^3,
\end{equation}
where $\LamJeansfluc$ is the perturbation Jeans length. Denoting $\lambda_{J_0}$ as the Jeans length at the position $r_0$ where the density of the envelope is $\rho_0$, the ratio between the mass of the density fluctuation and its Jog's mass can be written as 
\begin{align}
  \frac{\Mfluct}{\MJog}  &=   \left[ 1 - \left(\frac{\alpha-1}{3-\alpha} \right) \frac{1}{\eta} \right]^{3/2} \  \left( \frac{\delta_r}{\LamJeansfluc}\right)^3\nonumber\\
  &=  \left[ \eta- \frac{\alpha-1}{3-\alpha}\right]^{3/2}\  \left( \frac{\delta_r}{\LamJeanscero}\right)^3\  \left( \frac{r_{\rm p}}{r_0}\right)^{-3\alpha/2} ,
  \label{eq:mass_ratio}
\end{align}
%
therefore, the density fluctuation will be unstable if this ratio is larger than unity.
%
%
From this condition, we can obtain a minimal density contrast $\eta_{\rm crit}$ below which the mass of the density fluctuation is not large enough to collapse because of the tidal effect of the envelope:
\begin{equation}
    \eta_{\rm crit} = \left(\frac{\delta_r}{\LamJeanscero}\right)^{-2} 
             \left(\frac{r_{\rm p}}{r_0} \right)^\alpha
             +\frac{\alpha-1}{3-\alpha} .
   \label{eq:etacrit}
\end{equation}
\\

{
Equation (\ref{eq:etacrit}) is quite instructive. It tells us what would be the necessary value $\eta_{\rm crit}$ for a density fluctuation of size $\delta_{\rm r}$ within a power-law density profile, to become gravitationally unstable if it occurs at a distance $r_{\rm p}$ from the centre of the density profile. In Fig.~\ref{fig:eta_rp} we plot the critical density contrast $\eta_{\rm crit}$, as a function of $r_{\rm p}/r_0$, assuming that the size of the density fluctuation $\delta_r$ equals the local Jeans length $\LamJeanscero$. For large $r_{\rm p}/r_0 > 1$, the amplitude of the density fluctuation scales as $\eta_{\rm crit}\propto (r_{\rm p}/r_0)^\alpha$. For small $r_{\rm p}/r_0$ the amplitude of the density fluctuation becomes $\eta_{\rm crit}\leq 1$, and it is flat for $\alpha >1$, linear for $\alpha=1$, and decreases rapidly for $\alpha < 1$. The reason for this behavior is simple: since we assumed for this plot that $\delta_r = \lambda_0$, the density fluctuation has the Jeans length of the density at $r_0$. Thus, there is not much of a need for any density jump to become gravitationally unstable in the cases of $\alpha\ge1$, and due to their compression tides, shallower profiles even contribute to achieving the Jog mass.  \\

\begin{figure}
  \includegraphics[width=\columnwidth]{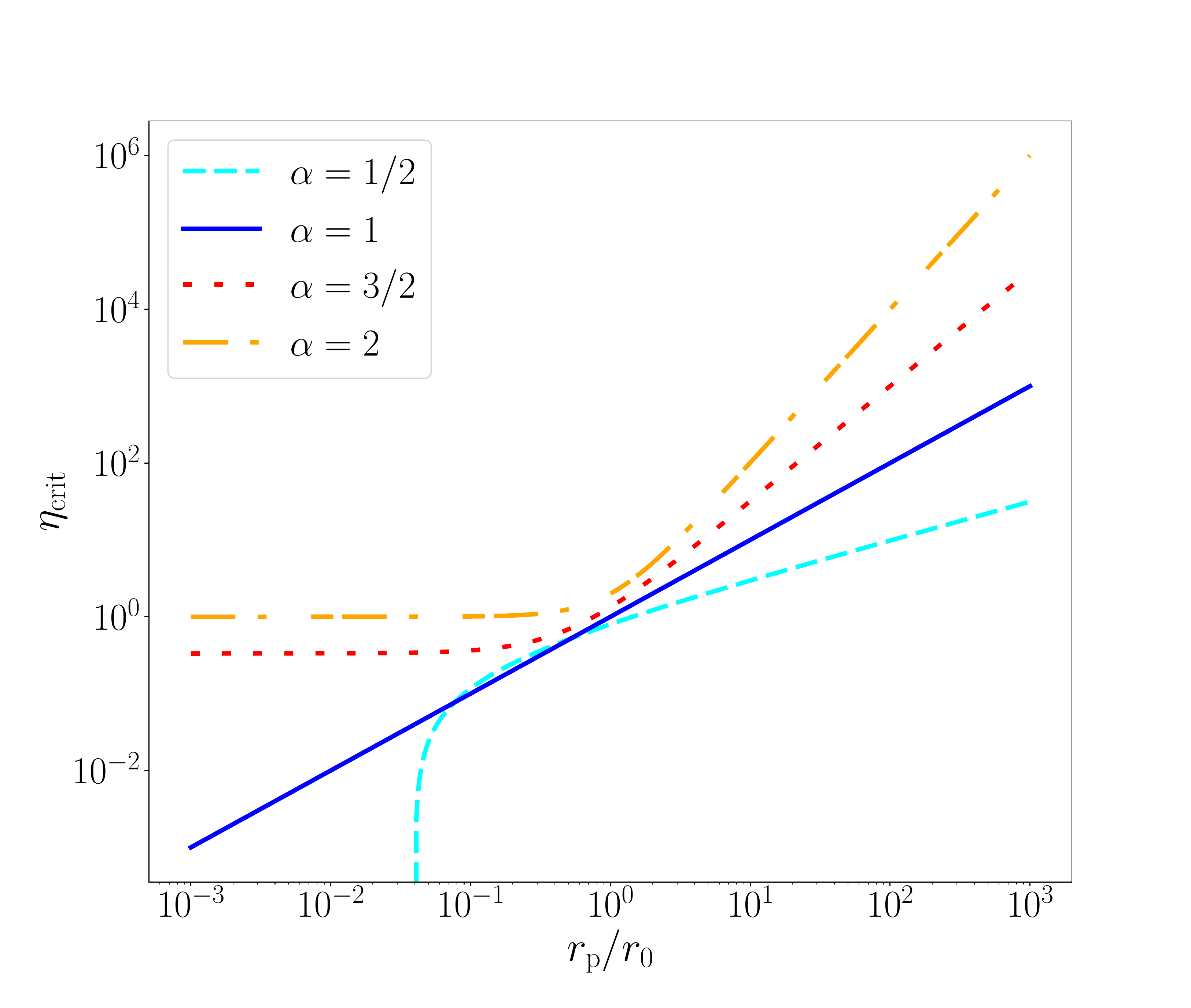}
   \caption{{ $\eta_{\rm crit}$ vs  $r_{\rm p}$. The solid, dotted, dashed, and dot-dashed lines represent the value of $\eta_{\rm crit}$ for $\alpha=$1/2, 1, 3/2 and 2, respectively.}
    \label{fig:eta_rp}}
\end{figure}

%
%
%
%
%
%
%

It is interesting to compare this figure to Fig.~10 in LH18. In this case, our $r_0$ can be taken as the Larson radius $r_{\rm L}$ in their analysis. The behavior of $\eta_{\rm crit}$ at large distances, $r_{\rm p} >> r_{\rm L}$ is similar to our case with $\alpha=2$: the larger the distance, the larger the density contrast necessary to produce a gravitationally unstable density fluctuation. However, while the exponent we found for this behavior is exactly $\alpha=2$, the value that one can infer from Fig.~10 in LH18 is about $\sim$1.8. \\

For small distances, $r_{\rm p} \ll r_{\rm L}$ the behavior is also similar, although again, not identical. While in our $\alpha=2$ case, the amplitude of $\eta$ becomes flat for small $r_{\rm p}/r_{\rm L}$, in the case of LH18 analysis the amplitude grows as $r_{\rm p}/r_{\rm L}$ decreases. The inferred behavior from their fig.~10 is $\eta_{\rm crit}\propto (r_{\rm p}/r_{\rm L})^{-0.9}$. This difference in behavior is consequence of the fact that at small $r_p$, tides from the Larson core itself become relevant, making it harder to produce collapse. Indeed, if we had included  the Larson core in our tidal analysis, the corresponding  dependency would be $\eta\propto (r_{\rm p}/r_{\rm L})^{-1}$ for $r< r_{\rm p}$. \\

Finally, we want to stress that the fluctuation within an isothermal density profile is not very realistic, since the Jeans length at any radius $r$ of an isothermal sphere with density amplitude $A=1$ is of the order of $\LamJeanscero\sim r$, while for an arbitrary $A$, the Jeans length will be $\LamJeanscero\sim r_0/\sqrt{A}$. This means that, unless $A$ is large, the size of a fluctuation within an isothermal sphere has to be comparable to the size of the sphere itself, in order to collapse. 
}
\subsection{Neglecting Larson core tidal effects}\label{sec:larson}
As we commented out at the beginning of \S\ref{sec:stab_jog_env}, our analysis does not include the contribution from the Larson core to the tidal streams.  In order to understand the relevance of a central Larson core on the stability of a density fluctuation in the presence of envelope tides, we have to compare the effect of the core in eq.~(\ref{eq:jog_mass}) with the effect of the envelope on the same equation. The effective density of the Larson core, given by the first component of $\tau^{\rm c}$ in eq.~(\ref{eq:marea_centro}), is
\begin{equation}
  \rhoefflarson (r)= \frac{3}{8\pi^2} \  {\rholarson},
\end{equation}
where $\rholarson$ is the density that will have a sphere with total mass equal to the Larson core, $M_L$ and size $r$, i.e., 
\begin{equation}
  \rholarson (r)= \frac{4\pi}{3}\  \frac{M_L}{r^3}.
\end{equation}
The ratio between the Larson's effective density and the density of the fluctuation is then given by
\begin{equation}\label{eq:cociente_densidad_equivalente_larson}
\frac{\rhoefflarson}{\rhofluct} = \frac{M_L}{2\pi\eta} \frac{r^{\alpha -3}}{\rho_0 r_0^\alpha},
\end{equation}
which for $\alpha<3$ decreases with $r$. In contrast, the ratio between the effective density of the envelope and the fluctuation density (see eq.~(\ref{eq:cociente_densidad_equivalente})) does not depend on $r$. \\

\begin{figure}
 \includegraphics[width=\columnwidth]{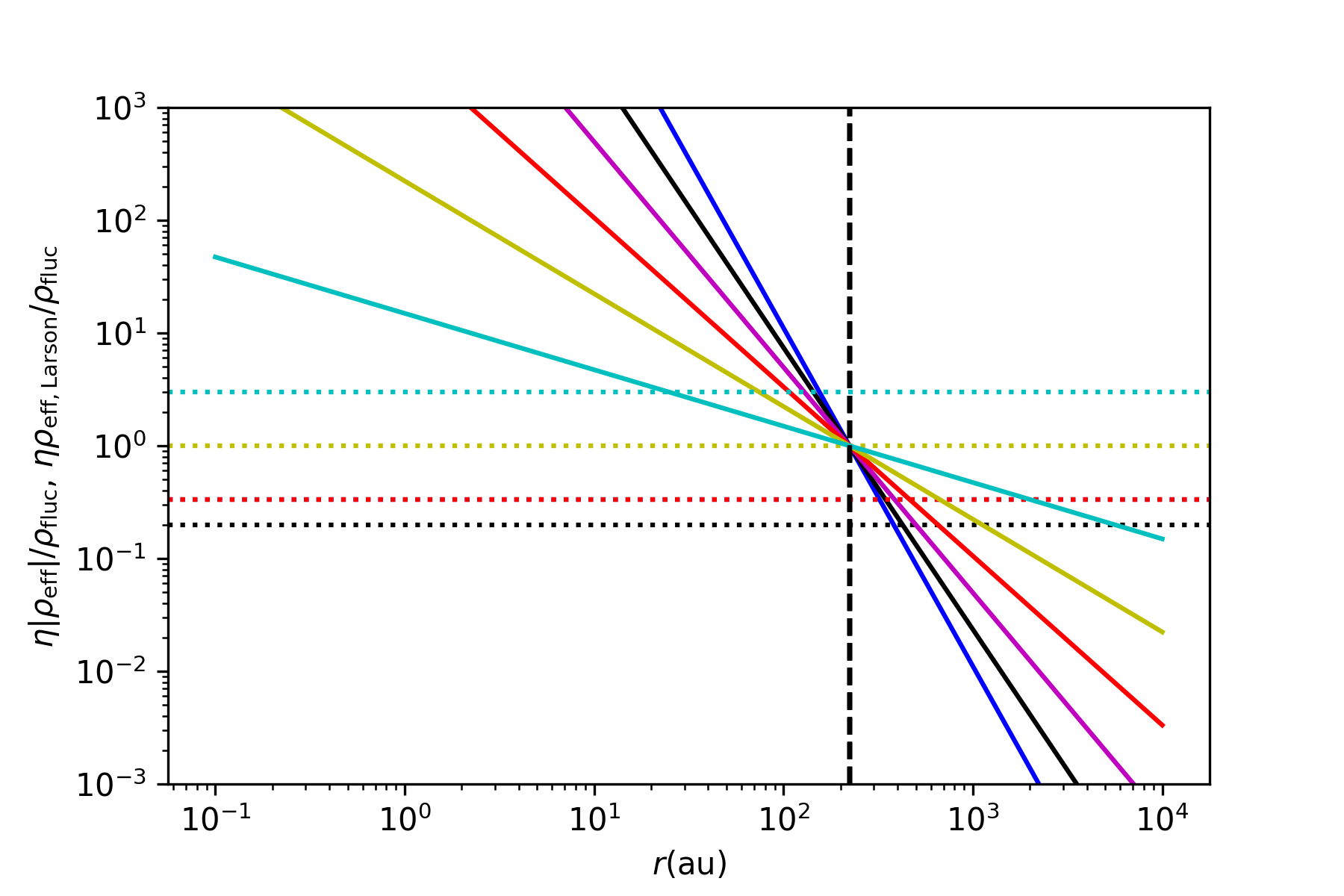} \\
 \caption{$\eta{\rhoefflarson}/{\rhofluct}$ (\textit{continuous lines}) and $\eta|{\rhoeff}/{\rhofluct}|$ (\textit{dotted lines}) as a function of $r$ for  $r_0=223$~AU and $\rho_0=\rhoefflarson(r_0)$. 
 The cyan, olive, red, purple, black and blue lines
denote the cases of $\alpha=$2.5, 2, 1.5, 1, 0.5 and 0, respectively. The vertical line is placed at $r=r_0$. }
    \label{fig:rhoefflarson_rhoeff1}
\end{figure}

In order to compare the tidal effects given by eqs.~(\ref{eq:cociente_densidad_equivalente}) and (\ref{eq:cociente_densidad_equivalente_larson}), in Fig. \ref{fig:rhoefflarson_rhoeff1} we plot the ratio $\eta\ {\rhoefflarson}/{\rhofluct}$  (solid lines), and $\eta\ |\ \rho_{\rm eff,env}/\rhofluct\ |$ (dotted lines) as a function of $r$. We have normalized the profiles such that $\rho_0 = \rhoefflarson$ at $r_0=223$~AU, which , for $c_{\rm s} = 0.2$~km s$^{-1}$, corresponds to the Larson radius $r_{\rm L} = GM_{\rm L}/{2 c_{\rm s}^2}$ of a 0.02~\Msun\ protostar at $T=11.4$~K, for a mean molecular mass $\mu = 2.36$.  The cyan, olive, red, purple, black and blue lines denote the cases $\alpha=$2.5, 2, 1.5, 1, 0.5 and 0, respectively\footnote{Note that the figure does not include the purple and the blue dotted lines. The first one corresponds to $\alpha=1$, in which case, the effective density of the envelope is zero, i.e., there are no tides from the envelope,  implying that the tidal effect of the Larson core is more relevant. The second one overlaps with the red dotted line due to the use of absolute values.}. Note that in this figure, we have used the absolute values, since the cases with $\alpha<1$ have negative effective densities.\\

From this figure it can be seen that, for a given $\alpha$ (denoted by lines with equal color), tides from the Larson core are more relevant wherever the solid line shows larger values than the corresponding dotted line. For instance, for steep envelope profiles (e.g., cyan line, $\alpha=$2.5), the tides from the Larson core dominate over the those of the envelope only for very small distances from the protostar ($\lesssim$30~AU). In contrast, in shallower profiles (e.g., red line, $\alpha=1.5$) the tides from the Larson core are relevant at scales of 500~AU. For $\alpha=1$, as we commented above, the envelope has no tides, and thus, those are always dominated by the Larson core. For even shallower profiles ($\alpha=$0.5, 0, black and blue lines, respectively) the tides of the Larson core can dominate over the tides of the envelope over scales as large as 1000~AU, although those tides are compressive, rather than disruptive. Note that all the distances just mentioned change with the normalization choice. In particular, for a fixed value of $\alpha$, higher (lower) values of $\rho_0$  reduce (increase) the radius at which the envelope tides become dominant over the central core tides.\\





An additional point requires to be emphasized: 
the fact that the Larson core can dominate the tides over the envelope does not mean that the Larson core contribution is necessarily relevant. Indeed, the relevance of any tide
depends on the ratio $|\rhoeff| /\rhofluct$. 
In particular,
the relevance of the Larson core tides decreases with $r$. For the normalization we used, all the density profiles imply $\rhofluct(r_L) =\eta\rhoefflarson$. Therefore, tides from the Larson core become progressively less relevant than the fluctuation self-gravity as $r>r_L$.


    
 

%
\section{Applications}
\label{sec:applications}

Equation (\ref{eq:mass_ratio}) can be renormalized, such that an observer can infer whether a density fluctuation within a molecular cloud will collapse or not in an easy way. 
As a good approximation, molecular clouds are isothermal, with temperatures of the order of 10~K. While they have a wide variety of scales, one can argue that MCs are typically tens of pc large, and are highly structured \citep[e.g., ][and references therein]{Ballesteros-Paredes+20}. Within them, cores with densities of the order of $\sim$~$10^{4}$~cm\alamenos3, and sizes of the order of $\sim$~0.1-0.5~pc are observed. Furthermore, those cores are often observed to have fragmentation \citep[e.g., ][]{Palau14,Beuther18}, with sizes of the order of $\delta_r\leq0.01-0.05$~pc. Taking into account the typical parameters measured for 
the density fluctuations or fragments\footnote{Previous observational works use the term `fragment' to what we have called density fluctuation, in reference to compact objects embedded in large-scale massive cores or clumps (called here `envelopes') undergoing fragmentation.
} embedded in large-scale cores,
eq.~(\ref{eq:mass_ratio}) can be rewritten as:
%

\begin{equation}
\begin{split}
 \frac{M_{\delta \rm r}}{M_{\rm Jog}} = \  2 \times 10^{-3} 
 \left( \frac{c_{\rm s}}{0.2~\si{\kilo\meter\per\second}} \right)^{-3} 
 &
 \left( \frac{n_0}{2\times10^{4}~\si{\per\cubic\centi\meter}} \right)^{3/2}  
 \left(\frac{r_{\rm p}}{r_0}\right)^{-3\alpha/2} 
 \\
 & \times
 \left( \frac{\delta_r}{0.02~\si{\parsec}} \right)^{3}
 \left( {\eta} - \frac{\alpha-1}{3-\alpha} \right)^{3/2},
\label{eq:criterio:colapso}
\end{split}
\end{equation}
where $n_0$ is the density of the large-scale core at the position of the fragment (i.e., $r_\mathrm{c}=r_0$) and $\delta_r$ is the fragment diameter. In this equation,  a mean molecular mass of $\mu = 2.36$ is adopted. Notice that we leave $r_0$ and $r_p$ without renormalization, since they can be written in the same units. We can now work with this equation in specific cases to evaluate whether observed density fluctuations are or not prone to collapse.

\subsection{Compact fragments within massive dense envelopes}\label{sec:Aina}

\citet{Palau21} studied the fragmentation properties of 18 massive dense  clumps/cores using Submillimeter Array (SMA) data in the extended/very extended configuration and single-dish telescope data. The massive dense clumps/cores in the \citet{Palau21} sample, referred here as `envelopes', lie at 1.3--2.6 kpc from the Sun, have total masses ranging from 40 to 1100~\mo, and bolometric luminosities from 200 to $1.4\times10^5$~\lo. They are all undergoing intermediate/high-mass star formation at relatively early evolutionary stages, as half of them have $L/M$ smaller than $\sim 10$ (with the $L/M$ ratio being an indicator of evolutionary stage proposed by \citealt{Molinari+16}). The interferometric study of the sample allowed these authors to resolve and characterize the compact fragments embedded in the envelopes. The SMA data were complemented with single-dish observations that were used to infer the spectral energy distribution of each envelope and the radial intensity profiles at several wavelengths. This allowed to estimate the density and temperature structures (modeled as power-laws) of the envelopes harboring the fragments. Thus, the data presented by \citet{Palau21} is well-suited to compare to the present work, as both the properties of the envelope ($n_0$, $\alpha$, temperature structure) and the properties of the individual fragments ($c_{\rm s}$, $\delta_r$, $\rhofluct$, $\eta$) could be directly inferred. Table~\ref{tab:cores1} lists the 18 massive dense envelopes of \citet{Palau21}, including the density and temperature structure reported in that work, along with the number of fragments associated with the central part of each envelope. In Table~\ref{tab:fragments1}, the complete list of the 160 compact fragments associated with the 18 envelopes is presented, including the corresponding position of each fragment.


In order to compare the aforementioned dataset with the theoretical work presented in this paper, we assumed that each fragment corresponds to a density fluctuation within the envelope. First, the distance from the centre of the envelope to each of its embedded fragments was estimated. In the ideal case of an envelope harboring one single massive protostellar object (massive fragment) surrounded by low-mass fragments, the peak of the submillimeter emission observed with the single-dish, from which the density power-law was inferred, should coincide with the position of the most massive fragment observed with the interferometer. However, this was not always the case in \cite{Palau21} data, mainly due to the single-dish position uncertainty (typically of about 2--3 arcsec) and because in some cases multiple massive protostellar objects were present inside the large-scale envelope. 
Here we will assume that the massive dense envelope for which the density structure was modeled using single-dish data is dominated by the fragment with highest intensity in the interferometric 0.87 or 1 mm image, and we will adopt the position of such bright fragment as the centre of the large-scale envelope (see Appendix~\ref{app} for further details).
Therefore, 142 out of 160 fragments remained to study in this work, because the fragments taken as centres of the large-scale envelopes are not considered.
Once the distance from the centre of the large-scale envelope to the position of each fragment is calculated, the temperature of the envelope at the position of the fragment, $T_\mathrm{env}$ (Table~\ref{tab:fragments1}), can be inferred, using the temperature power-law modeled by \cite[given in Table~\ref{tab:cores1}]{Palau21}.

In addition, using the flux density at 0.87 or 1 mm and the diameter $\delta_r$ of each fragment  
(both taken from the 3$\sigma$ contour of the interferometric images), one can estimate the mass and density of the fluctuation without taking into account the contribution of the envelope (because the interferometre filters out the large-scale emission). The mass of each fragment, $M_\mathrm{\delta r}$, was estimated using the temperature of the envelope at the position of the fragment ($T_\mathrm{env}$, see previous paragraph), and the dust opacity from \cite{Ossenkopf_Henning94}. These quantities are listed in Table~\ref{tab:fragments1}. 

Finally, to estimate the ratio $M_\mathrm{\delta r}/M_\mathrm{Jog}$ following eq.~(\ref{eq:criterio:colapso}), the density of the fragment $n_\mathrm{fluct}$, the density of the envelope at the position of the fragment $n_\mathrm{env}$, the density contrast $\eta$, and the sound speed, are also required. First, $n_\mathrm{fluct}$ was obtained from $M_\mathrm{\delta r}$ and $\delta$ (see previous paragraph and Table~\ref{tab:fragments1}). Second, using the density power-law of the envelope (Table~\ref{tab:cores1}) and the distance of each fragment to the envelope centre (Table~\ref{tab:fragments1}), the density of the envelope at the position of each fragment, $n_\mathrm{env}$, was obtained. From these two quantities, the density contrast is directly obtained as $\eta=n_\mathrm{fluct}/n_\mathrm{env}$. Third, using $T_\mathrm{env}$ (Table~\ref{tab:fragments1}), the sound speed at the position of each fragment was calculated. With these quantities, all listed in Table~\ref{tab:fragments2}, the ratio $M_\mathrm{\delta r}/M_\mathrm{Jog}$ was calculated for each fragment.

\begin{figure}
  \includegraphics[width=\columnwidth]{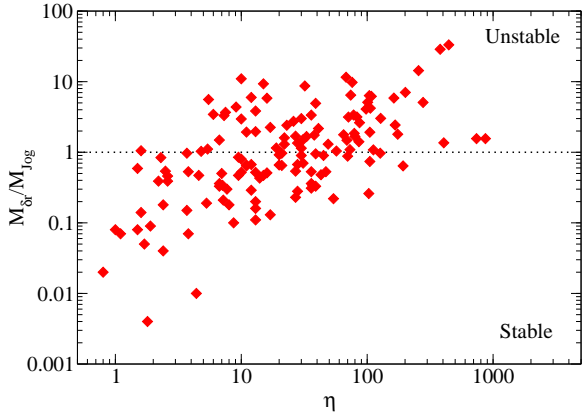}
   \caption{Ratio of the mass of the fluctuation (fragment) over the Jog mass for the 142 fragments studied in the sample of \citet{Palau21}. About half of the fragments have ratios above 1, indicating that they are gravitationally unstable when the tides of their parent clump are taken into account.
   }
    \label{fig:Aina}
\end{figure}

In Fig.~\ref{fig:Aina}, we present the ratio $M_\mathrm{\delta r}/M_\mathrm{Jog}$ vs. the density contrast $\eta$, where an increasing trend is appreciated, as expected from equation (\ref{eq:criterio:colapso}). As seen from this figure, about half of the fragments ($\sim49$\%) have ratios above 1, indicating that they are unstable and undergoing collapse. Using the density of the fragment and the temperature of the envelope at the position of the fragment to calculate the Jeans mass, we found that the Jog masses are about 30\% larger than their corresponding Jeans masses, and that the percentage of unstable fragments using the Jeans mass (with the density of the fragment) is larger than that obtained using the Jog mass (59\%\ vs 49\%). The fact that the Jog masses are in most cases larger than the Jeans masses (calculated using the density of the fragment) stems from the fact that the density power-law indices for all the cores of the \cite{Palau21} sample are always larger than 1 and therefore tides are disruptive in all the cores according to eq.~(\ref{eq:cociente_jog_jeans}).

\subsection{Can the observed fragments be formed by turbulence?}\label{sec:formationfragments}

Supersonic shocks in a turbulent environment produce density fluctuations, and their amplitude scales as $\eta\equiv\delta\rho/\rho={\cal{M}}^2$ \citep[e.g., ][]{Vazquez-Semadeni94, Ballesteros-Paredes+07}, with ${\cal M}$ being the turbulent Mach number of the environment. Thus, it is natural to ask whether the observed density fluctuations within the envelope could have been produced by turbulence. \\ 

We envise two possibilities for turbulent motions: global interstellar turbulence and collapse-induced turbulence. In the former case, turbulent motions result from a continuous cascade of energy. Indeed, turbulent supersonic energy is introduced at the largest scales by, e.g., accretion from extragalactic gas \citep{Klessen_Hennebelle10}; the global action of supernova remnants \citep[e.g., ][]{Padoan+16, Walch+15, Ibanez-Mejia+16}; spiral density waves \citep{Martos_Cox98, Wada+02, Gomez_Cox02, Kevlahan_Pudritz09}; tidal forces from molecular cloud complexes \citep{Mao+20, Ramirez-Galeano+22}, etc. For a review of these mechanisms, see \citet{Klessen_Glover16}. Such energy injection is transferred to smaller scales until viscosity dissipates it. In this process, turbulent motions transit from being supersonic at large scales to subsonic at smaller scales. Statistically speaking, this transition is found to occur typically at scales of 0.1~pc \citep[e.g.,][]{Myers83, Heyer_Brunt04}. \\

In the present work, we wonder about the conditions in which turbulence produces density fluctuations that can collapse on scales substantially smaller than 0.1~pc. For instance, the fragments studied in \S\ref{sec:Aina} have sizes between 0.002 and 0.02~pc, and substantial density contrasts spaning three orders of magnitude (see Fig.\ref{fig:Aina}). Thus, it is unlikely that subsonic interstellar turbulence has produced them. \\

The other possibility for turbulence to produce those fragments is to be generated locally by the very collapse of the envelope, as suggested by \citet{Lee_Hennebelle18b}. Indeed, according to \citet{Guerrero_Gamboa+20}, $\sim$40\% of the infall velocity can be converted into turbulence during collapse \citep[a value in agreement with the estimations by][]{Lee_Hennebelle18b}, i.e., 
%
\begin{equation}
\effturb =  \frac{\sigmaturb}{\upsilon_{\rm ff}} \sim 0.4
\end{equation}
where $\epsilon_{\rm ff-turb}$ is the efficiency factor in converting infalling motions into turbulent ones, $\sigma_{\rm ff-turb}$ is the velocity dispersion of the turbulent motions (originated by the free-fall), and $v_{\rm ff}$ is the free-fall velocity of a particle at a position $r$. Thus, the collapse of a mass $\Menv$ from infinity to size $r_p$ where the density fluctuation will occur can develop free-fall velocities up to 
%
\begin{equation}
\upsilon_{\rm ff} = \sqrt{\frac{2 G \Menv(r_{\mathrm{p}})}{r_{\mathrm{p}}}},
\end{equation}
which in turn will develop turbulence with Mach numbers of the order of 
\begin{equation}
\Machffturb = \epsilon_{\rm ff-turb} \ \bigg( \frac{\upsilon_{\rm ff}}{c_s}  \bigg)   = \frac{\epsilon_{\rm ff-turb}}{c_s} \sqrt{\frac{2 G \Menv(r_{\mathrm{p}})}{r_\mathrm{p}}}. \\
\label{eq:Machffturb}
\end{equation}

\begin{figure}
    \includegraphics[width=\columnwidth]{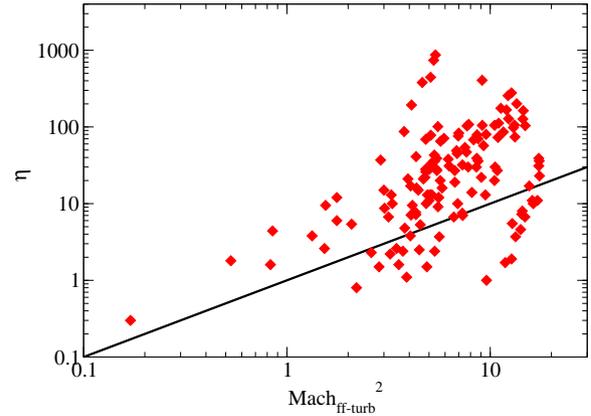}
   \caption{$\eta$ vs. $\mathcal{M}_{\rm ff-turb}^2$ for the 142 fragments studied in the sample of \citep{Palau21}. The solid black line corresponds to the one-to-one relation. Most of the fragments present $\eta \gg \mathcal{M}_{\rm ff-turb}^2$, suggesting that the turbulence generated by the collapse of the massive dense core is not enough to account for the density of the observed fragments.}
    \label{fig:Aina3}
\end{figure}

Now, the question is whether the Mach number of turbulent motions induced by the collapse of the envelope with mass $\Menv$, is similar to the value of the density contrast of the observed fragments, i.e., whether $\Machffturb^2 \geq \eta$. If so,  the turbulence generated by the collapse of the envelope could be the responsible for its own fragmentation. In Fig.~\ref{fig:Aina3} we plot these two quantities for the observational data presented in \S\ref{sec:Aina}. The solid line denotes the locus $\Machffturb^2 = \eta$, and thus, fragments above this line have density contrasts larger than the square of the turbulent Mach number. 
As can be seen in the figure,  most of the fragments have $\eta\ \gg \Machffturb^2$, by up to 2 orders of magnitude. This could be in part because some of the fragments are already undergoing collapse and therefore their actual density contrast does not directly relate to the Mach number, while another fraction of the fragments might not be gravitationally unstable and their density contrast is just the result from the primordial density fluctuations in the envelope. \\

We conclude, thus, that the observed fragments are not likely the result of the global interstellar turbulence acting at the moment of collapse of the envelope, and most of them will not be generated during the collapse of the envelope. This implies that such fragments or fluctuations were probably generated by the turbulence developed at the time of the cloud formation, and therefore were already present in the parental envelope before its collapse, when the envelope was shallower.\\
 
\subsection{Can a particular molecular cloud core undergo turbulent fragmentation in the future? The case of B68}

Using eq.~(\ref{eq:criterio:colapso}), we can also make assumptions on the different quantities and ask  what turbulent Mach number ${\cal M}^2=\eta$ can give rise to gravitationally unstable density fluctuations.
i.e., $ \Mfluct/\MJog>1$. \\

Let us take, for instance, the case of Barnard 68. This is an isolated molecular cloud core with a radius $R = 0.06~\si{\parsec}$, mass $M= 2.1~\text{M}_{\odot}$ \citep{Alves+01}, which gives a mean density of $n_0 \sim 4 \times 10^{4}~\si{\per\cubic\centi\meter}$. It has a temperature of $T=16~\si{\kelvin}$, implying a sound speed of $c_{\rm s} \sim 0.24~\si{\kilo\meter\per\second}$  \citep{Bourke+95}. We can also assume, furthermore, that we want to generate a fragment with size $\delta_r\sim$0.01~pc\footnote{Producing smaller fragments will be substantially more difficult since the dependency of the ratio $M_{\delta_r}/\MJog$ goes as $\delta_r^3$, see eq.~(\ref{eq:criterio:colapso}).} ($\sim$2000~AU). Assuming that $r_p\sim r_0$, which for this core is $\sim 6000$~AU, from eq.~(\ref{eq:etacrit}) it follows that the minimal required Mach number is 
\begin{equation}
    {\cal{M}} \sim  \sqrt{\frac{\alpha-1}{3-\alpha} + 176}.
\end{equation}
Although the value of $\alpha$ is uncertain, it is clear from this equation that the 
necessary Mach number for producing a turbulent density fluctuation that achieves enough mass to collapse is quite large (of $\sim13$ for typical values of $\alpha \sim 1$--2), making it unlikely that further turbulent fragmentation can occur in B68.

\section{Discussion and conclusions}\label{sec:discussion}

In the present work, we have analyzed the role of tidal forces over a density fluctuation occurring in a spherical envelope. On the one hand, the relevant components producing tides are the presence of a protostar (Larson core) and, on the other hand, the envelope where the density fluctuation occurs. \\

In order to estimate the relevance of the tides, we made use of the formalism presented by \citet{Jog13}. This formalism is an extension of the Jeans analysis where an external potential is present. In addition, our analysis has the advantages of including the thermal support role and allowing the power-law density profile to vary. This is relevant because while it is true that collapsing objects develop a power law with index of $\alpha = -2$, in general star-forming regions may present a variety of density profiles. \\

In our development, we computed new collapsing conditions. While typically, one considers the Jeans mass as the mass to be exceeded for a fluctuation to collapse, in the presence of tides, the equivalent is the Jog mass. As expected, both masses tend to be similar for large density fluctuations because tides become less relevant when self-gravity becomes more important. However, for smaller density fluctuations, the effect of tides can be relevant, but that depends on the density profile of the envelope. 
 In addition, our analysis also suggests that the clustered mode of star formation may originate from massive cores or envelopes having relatively flat profiles ($\alpha < 1.5$), while the isolated mode should be produced when the core has steeper ($\alpha\geq 1.5$) density profiles.\\

We also found that the Larson core has little relevance unless we are dealing with scales smaller than the Larson radius, which is of the order of 200~AU. This implies that the Larson core has only a major effect if we analyze, for instance, the stability of a density fluctuation at very small scales, such as those of a protoplanetary disk. \\

We finally applied our formalism to estimate the stability of 142 fragments observed in regions of high-mass star formation reported in previous works. These fragments, embedded in larger clumps of gas for which the density and temperature have been estimated, 
 present Jog masses larger (by 30\%) than the Jeans mass (calculated using the fragment density). This is because all the observed massive dense cores/envelopes have density power-law indices larger than 1, producing disruptive tides.\\

We also estimated the required turbulent Mach numbers for producing collapsing density fluctuations in typical envelopes. We found that the fragments observed have been hardly produced by the interstellar turbulence acting at the moment of collapse because, at the scales of these fragments, interstellar turbulence is already subsonic and will hardly produce fragments with the density contrast and size of the observed fragments. Similarly, we have shown that the Mach number required to generate the observed density fluctuations are substantially larger than the Mach numbers that the parent core can generate in transferring collapse energy into turbulent energy. This result is relevant because it has been proposed that some levels of turbulence can be generated by the collapse of molecular clouds \citep{Guerrero_Gamboa+20}, and that such collapse may generate density fluctuations that can be able to collapse and compete for the available mass reservoir in molecular cloud cores \citep{Lee_Hennebelle18b, Colman_Teyssier20}. We found that, in practice, this is an unlikely mechanism. \\

Thus, it is relevant to consider the tidal effects on the stability of  cores, clumps and clouds, and we suggest that the currently observed fragmentation should have a more primordial origin rather than being produced in the present by the interstellar turbulence or the very collapse of the protostellar envelopes themselves.


%

\section*{Acknowledgements}

We are greatly grateful to Chanda J. Jog, and to an anonymous referee for their insightful and encouraging comments.

RZ-M acknowledges scholarship from UNAM-DGAPA-PAPIIT grant number {\tt IN-111-219} during 2021, {\tt IN-118-119} during 2022; and from CONACYT project number {\tt INFR-2015-253691}. He also acknowledges to UNAM-DGAPA PAPIIT grant number {\tt AG-101-723} for travel support in 2023.
JB-P  acknowledges UNAM-DGAPA-PAPIIT support through grant number {\tt IN-111-219}, CONACYT, through grant number {\tt 86372}, and to the Paris-Saclay University’s Institute Pascal for the invitation to ‘The Self-Organized Star Formation Process’ meeting, in which invaluable discussions with the participants lead to the development of the idea behind this work. 
AG acknowledges UNAM-DGAPA-PAPIIT support through grant number {\tt IN-115-623}.
AP acknowledges financial support from the UNAM-PAPIIT {\tt IN-111-421} and {\tt IG-100-223} grants, the Sistema Nacional de Investigadores of CONACyT, and from the CONACyT project number {\tt 86372} of the `Ciencia de Frontera 2019’ program, entitled `Citlalc\'oatl: A multiscale study at the new frontier of the formation and early evolution of stars and planetary systems’, M\'exico.

\section*{Data Availability}

The data underlying this paper will be shared on reasonable request to the corresponding author.
 



\bibliographystyle{mnras}
\bibliography{mnras_template.bib}




\appendix
\section{Properties of the massive dense cores and their embedded compact fragments used in this work}\label{app}

In this Appendix we present the data used to make Figs.~\ref{fig:Aina} and \ref{fig:Aina3}. Table~\ref{tab:cores1} lists the sample of massive dense cores (corresponding to `envelopes', following the nomenclature of this work) studied by \citet{Palau21} at scales of about 0.15 pc of diameter. The properties of these massive dense envelopes were obtained using single-dish telescopes. These envelopes were modeled assuming that the density and temperature follow power-laws, as also assumed in this work. Regarding the centre of the envelope listed in Table~\ref{tab:cores1}, this was chosen to be the position of the fragment with highest intensity at 0.87 or $\sim1$~mm, identified using interferometres by \citet{Palau21}\footnote{ This applies to all the cases except for G34.4.1, and CygX-N3. In these two cases, the centre of the envelope was taken as the position of the second fragment with highest intensity, because in these two cases the second fragment was much closer to the JCMT peak at 450~$\mu$m and its intensity was very similar (about 80\%) to the intensity of the strongest fragment.}. A visual inspection was done to confirm that such a fragment lies within 2--4~arcsec of the peak of the submillimetre core as traced with the James Clerk Maxwell Telescope (JCMT) at 450~$\mu$m \citep{DiFrancesco08} in most of the cases\footnote{Only for the cases of NGC\,6334I, NGC\,6334In, and G35.2N the separation between the strongest fragment and the JCMT peak was 6--8 arcsec, above the typical positional uncertainty of the single-dish of $\sim2$ arcsec. In these two cases the spatial distribution and intensities of the fragments surrounding the strongest one is the responsible for such a larger separation.}. Table~\ref{tab:cores1} also gives the number of millimetre compact fragments identified within each massive envelope by \citep{Palau21}, which makes a total of 160 fragments.

In Table~\ref{tab:fragments1}, the entire sample of 160 fragments associated with the massive dense envelopes of Table~\ref{tab:cores1} is presented. For each fragment, the table lists its coordinates, distance to the envelope centre (see above), temperature of the envelope at the position of the fragment, flux density, frequency of the interferometric observations used to identify the fragment, adopted dust opacity, estimated mass, and size.

Finally, in Table~\ref{tab:fragments2}, the specific calculations performed to apply the theoretical results of this work to the sample of 142 fragments are reported (the fragments taken as reference are not considered). Assuming that each fragment corresponds to a density fluctuation within the envelope, Table~\ref{tab:fragments2} reports the density of the fluctuation (fragment), the density of the envelope at the position of the fluctuation, the amplitude of the fluctuation $\eta$, the sound speed at the position of the fluctuation, the ratio of the mass of the fluctuation over the Jog mass (given by equation \ref{eq:criterio:colapso}), the Jog mass, the mass of the envelope enclosed within a radius $r_\mathrm{p}$ (the position of the fluctuation) and the Mach number of the turbulence generated by the gravitational collapse of the envelope (calculated following equation \ref{eq:Machffturb}). The tablenotes of the tables of this Appendix provide all the details of these calculations.

\begin{table*}
\caption{Properties of the sample of massive dense cores/envelopes studied in \citet{Palau21}.}
\label{tab:cores1}
\begin{tabular}{lcccccccc}
\hline
        &D\supa      &R.A.\supb      &Dec.\supb        &$T_\mathrm{1000}$\supc  &       &$\rho_\mathrm{1000}$\supc\\
Core/envelope    &(kpc)  &(J2000)   &(J2000)     &(K)    &$q$\supc    &(g\,cm$^{-3}$) &$\alpha$\supc &$N_\mathrm{mm}$\supd\\
\hline
W3IRS5          &1.95   &02:25:40.773   &+62:05:52.60   &260     &0.40 &$2.4\times10^{-17}$	&1.46   &4\\
W3H2O           &1.95   &02:27:03.852   &+61:52:25.03   &152     &0.38 &$1.5\times10^{-16}$ &1.90   &8\\
G192            &1.52   &05:58:13.536   &+16:31:58.19   &66      &0.37 &$4.1\times10^{-17}$ &2.13   &1\\
NGC\,6334V      &1.3    &17:19:57.373   &$-35$:57:53.24	&96      &0.32 &$1.3\times10^{-16}$	&1.89   &5\\
NGC\,6334A      &1.3    &17:20:17.856	&$-35$:54:43.75	&78	     &0.32 &$4.6\times10^{-17}$	&1.42   &16\\
NGC\,6334I      &1.3    &17:20:53.434   &$-35$:46:58.18 &111     &0.31 &$2.3\times10^{-16}$	&2.02   &7\\
NGC\,6334In     &1.3    &17:20:55.245   &$-35$:45:03.90	&98	     &0.33 &$8.8\times10^{-17}$	&1.46   &15\\
G34.4.0         &1.57   &18:53:18.005   &+01:25:25.51	&63	     &0.34 &$2.0\times10^{-16}$	&2.26   &5\\
G34.4.1         &1.57   &18:53:18.698   &+01:24:41.45	&63	     &0.37 &$8.2\times10^{-17}$	&1.76   &10\\
G35.2N          &2.19   &18:58:12.954   &+01:40:37.31	&90	     &0.34 &$4.4\times10^{-17}$	&1.53   &15\\
I20126          &1.64   &20:14:26.030	&+41:13:32.56	&86	     &0.34 &$8.4\times10^{-17}$	&2.21   &1\\
CygX-N3         &1.40   &20:35:34.410	&+42:20:07.00	&45	     &0.35 &$4.0\times10^{-17}$	&1.58   &6\\
W75N            &1.40   &20:38:36.497   &+42:37:33.50	&112     &0.33 &$1.4\times10^{-16}$	&1.99   &14\\
DR21OH          &1.40   &20:39:00.999	&+42:22:48.91	&73	     &0.36 &$3.1\times10^{-16}$	&1.98   &18\\
CygX-N48        &1.40   &20:39:01.340	&+42:22:04.76	&58	     &0.34 &$1.0\times10^{-16}$	&1.71   &12\\
CygX-N53        &1.40   &20:39:02.959   &+42:25:50.99	&45	     &0.36 &$9.7\times10^{-17}$ &1.76   &9\\
CygX-N63        &1.40   &20:40:05.378   &+41:32:13.04	&45	     &0.34 &$6.5\times10^{-17}$ &2.03   &2\\
NGC\,7538S      &2.65   &23:13:44.989	&+61:26:49.77	&93	     &0.35 &$1.3\times10^{-16}$	&1.72   &12\\
\hline
\end{tabular}
\begin{list}{}{}
\item[$^\mathrm{a}$] Distance to the source.
\item[$^\mathrm{b}$] Reference position taken as the centre of the massive dense envelopes and used to infer the properties of the fragments presented in Tables~\ref{tab:fragments1} and \ref{tab:fragments2}. This position corresponds to the position of the fragment with highest intensity detected at $\sim 1$mm with interferometres, also listed in Table~\ref{tab:fragments1}. The position of this fragment was confirmed to be near the centre of the envelope as traced with the James Clerk Maxwell Telescope at 450~$\mu$m \citep{DiFrancesco08} in most of the cases (see the Appendix text for more details).
\item[$^\mathrm{c}$] Density and temperature power-laws used to model the envelopes by \citet{Palau21}, as $\rho_{\rm env}(r) = \rho_\mathrm{1000} (r/r_\mathrm{1000})^{-\alpha}$ and $T_{\rm env}(r) = T_\mathrm{1000} (r/r_\mathrm{1000})^{-q}$. $\rho_\mathrm{1000}$ and $T_\mathrm{1000}$ correspond to the density and temperature at the reference radius taken as 1000 au.
\item[$^\mathrm{d}$] $N_\mathrm{mm}$ corresponds to the number of millimetre compact fragments embedded within each envelope, identified by \citet{Palau21} using interferometric data, and within a region of 0.15~pc of diameter (the common field of view for all the envelopes).
\end{list}
\end{table*}

\begin{table*}
\caption{Position, mass and size of the 160 compact fragments embedded in the massive dense envelopes of \citet{Palau21}.}
\label{tab:fragments1}
\begin{tabular}{lccccccccccc}
\hline
&Fragment R.A. & Fragment Dec. &$r_\mathrm{p}$\supa &$r_\mathrm{p}$\supa &$T_{\rm env}$\supb &S$_\nu$\supc &Freq.\supc &Opacity\supc &$M_\mathrm{\delta r}$\supc &$\delta r/2$\supd &$\delta r/2$\supd \\         
Fragment   &(J2000) &(J2000) &(arcsec) &(au) &(K)   &(mJy)&(GHz) &($\si{\square\centi \meter\per\gram}$) &(\mo) &(arcsec) &(au)\\
\hline
W3IRS5-MM1	&02:25:40.523	&+62:05:50.22	&2.96	&5770	&129	&29	&345	&0.01751	&0.069	&0.73	&1420\\
W3IRS5-MM2	&02:25:40.682	&+62:05:52.04	&0.85	&1660	&212	&62	&345	&0.01751	&0.086	&0.80	&1560\\
W3IRS5-MM3\supe&02:25:40.773&+62:05:52.60	&0.00	&0	    &260	&80	&345	&0.01751	&0.090	&0.79	&1540\\
W3IRS5-MM4	&02:25:41.022	&+62:05:48.09	&4.84	&9440	&106	&28	&345	&0.01751	&0.082	&0.74	&1440\\
\hline
W3H2O-MM1	&02:27:03.359	&+61:52:30.70	&6.65	&12980	&57	    &14	&218	&0.00798	&0.41	&0.21	&410\\
W3H2O-MM2	&02:27:03.382	&+61:52:18.93	&6.94	&13540	&56	    &13	&218	&0.00798	&0.39	&0.20	&390\\
W3H2O-MM3\supe&02:27:03.852	&+61:52:25.03	&0.00	&0	    &152	&2000&218	&0.00798	&$-$	&$-$	&$-$\\
W3H2O-MM4	&02:27:03.983	&+61:52:26.69	&1.90	&3710	&92	    &14	&218	&0.00798	&0.26	&0.22	&430\\
W3H2O-MM5	&02:27:04.299	&+61:52:24.02	&3.32	&6470	&75	    &70	&218	&0.00798	&1.57	&0.43	&840\\
W3H2O-MM6	&02:27:04.374	&+61:52:26.40	&3.93	&7670	&70	    &11	&218	&0.00798	&0.27	&0.20	&400\\
W3H2O-MM7	&02:27:04.569	&+61:52:24.60	&5.08	&9910	&64	    &360&218	&0.00798	&9.62	&0.63	&1230\\
W3H2O-MM8	&02:27:04.724	&+61:52:24.67	&6.17	&12030	&59	    &381&218	&0.00798	&11.03	&0.71	&1380\\
\hline
G192-MM1	&05:58:13.536	&+16:31:58.19	&0.00	&0	    &66	    &547&345	&0.01751	&1.63	&1.16	&1770\\
\hline
N6334V-MM1	&17:19:57.080	&$-35$:57:53.46	&3.56	&4630	&59	&65	&345	&0.01751	&0.16	&0.72	&940\\
N6334V-MM2\supe&17:19:57.373&$-35$:57:53.24	&0.00	&0	    &96	&1210&345	&0.01751	&1.74	&1.44	&1870\\
N6334V-MM3	&17:19:57.545	&$-35$:57:53.32	&2.09	&2720	&70	&604&345	&0.01751	&1.24	&1.23	&1600\\
N6334V-MM4	&17:19:57.619	&$-35$:57:48.52	&5.58	&7260	&51	&41	&345	&0.01751	&0.12	&0.58	&750\\
N6334V-MM5	&17:19:57.817	&$-35$:57:53.02	&5.39	&7010	&51	&229&345	&0.01751	&0.66	&1.08	&1400\\
\hline
N6334A-MM1	&17:20:17.749	&$-35$:54:41.78	&2.36	&3070	&54	&170	&345	&0.01751	&0.46	&0.64	&830\\
N6334A-MM2	&17:20:17.773	&$-35$:54:43.58	&1.02	&1330	&71	&1060	&345	&0.01751	&2.12	&1.41	&1840\\
N6334A-MM3	&17:20:17.791	&$-35$:54:41.86	&2.05	&2660	&57	&139	&345	&0.01751	&0.36	&0.60	&780\\
N6334A-MM4\supe&17:20:17.856&$-35$:54:43.75	&0.00	&0	    &78	&672	&345	&0.01751	&1.21	&1.16	&1510\\
N6334A-MM5	&17:20:17.949	&$-35$:54:39.60	&4.30	&5590	&45	&29	    &345	&0.01751	&0.10	&0.57	&740\\
N6334A-MM6	&17:20:18.036	&$-35$:54:41.94	&2.84	&3690	&51	&430	&345	&0.01751	&1.25	&1.27	&1660\\
N6334A-MM7	&17:20:18.201	&$-35$:54:32.43	&12.07	&15690	&32	&52	    &345	&0.01751	&0.26	&0.73	&950\\
N6334A-MM8	&17:20:18.464	&$-35$:54:41.44	&7.74	&10060	&37	&211	&345	&0.01751	&0.90	&1.00	&1310\\
N6334A-MM9	&17:20:18.664	&$-35$:54:38.88	&10.96	&14250	&33	&236	&345	&0.01751	&1.16	&0.95	&1230\\
N6334A-MM10	&17:20:18.705	&$-35$:54:40.27	&10.89	&14150	&33	&764	&345	&0.01751	&3.74	&1.44	&1870\\
N6334A-MM11	&17:20:18.819	&$-35$:54:39.55	&12.43	&16160	&32	&396	&345	&0.01751	&2.05	&1.11	&1450\\
N6334A-MM12	&17:20:18.885	&$-35$:54:37.88	&13.81	&17960	&31	&136	&345	&0.01751	&0.73	&0.92	&1200\\
N6334A-MM13	&17:20:19.026	&$-35$:54:37.71	&15.45	&20080	&30	&119	&345	&0.01751	&0.67	&0.94	&1220\\
N6334A-MM14	&17:20:19.292	&$-35$:54:37.25	&18.62	&24210	&28	&84	    &345	&0.01751	&0.51	&0.72	&940\\
N6334A-MM15	&17:20:19.313	&$-35$:54:34.48	&19.98	&25980	&28	&114	&345	&0.01751	&0.72	&0.85	&1110\\
N6334A-MM16	&17:20:19.399	&$-35$:54:35.82	&20.36	&26460	&27	&541	&345	&0.01751	&3.44	&1.25	&1630\\
\hline
N6334I-MM1	&17:20:52.673	&$-35$:47:02.54	&10.23	&13300	&50	&49    &345	&0.01751	&0.15	&0.42	&550\\
N6334I-MM2	&17:20:52.914	&$-35$:46:57.93	&6.33	&8230	&58	&79	   &345	&0.01751	&0.20	&0.53	&690\\
N6334I-MM3	&17:20:53.093	&$-35$:47:01.12	&5.08	&6610	&62	&215   &345	&0.01751	&0.50	&0.59	&770\\
N6334I-MM4	&17:20:53.104	&$-35$:47:03.88	&6.97	&9060	&56	&242   &345	&0.01751	&0.64	&0.70	&910\\
N6334I-MM5	&17:20:53.124	&$-35$:47:06.82	&9.43	&12250	&51	&57	   &345	&0.01751	&0.17	&0.51	&670\\
N6334I-MM6	&17:20:53.193	&$-35$:46:59.82	&3.36	&4370	&70	&2970  &345	&0.01751	&6.02	&1.40	&1820\\
N6334I-MM7\supe&17:20:53.434&$-35$:46:58.18	&0.00	&0	    &111&7080  &345	&0.01751	&8.69	&1.62	&2100\\
\hline
N6334In-MM1	&17:20:54.133	&$-35$:45:13.46	&16.56	&21540	&36	&72	    &345	&0.01751	&0.32	&0.55	&720\\
N6334In-MM2	&17:20:54.557	&$-35$:45:18.32	&16.67	&21680	&36	&384	&345	&0.01751	&1.74	&0.89	&1150\\
N6334In-MM3	&17:20:54.560	&$-35$:45:16.18	&14.84	&19290	&37	&350	&345	&0.01751	&1.51	&0.92	&1200\\
N6334In-MM4	&17:20:54.643	&$-35$:45:17.32	&15.29	&19880	&37	&421	&345	&0.01751	&1.84	&0.81	&1050\\
N6334In-MM5	&17:20:54.694	&$-35$:45:08.51	&8.14	&10580	&45	&156	&345	&0.01751	&0.53	&0.74	&960\\
N6334In-MM6	&17:20:54.719	&$-35$:45:15.05	&12.86	&16710	&39	&133	&345	&0.01751	&0.54	&0.66	&850\\
N6334In-MM7	&17:20:54.739	&$-35$:45:00.92	&6.84	&8890	&48	&261	&345	&0.01751	&0.83	&0.92	&1200\\
N6334In-MM8	&17:20:54.774	&$-35$:45:14.42	&11.98	&15570	&40	&114	&345	&0.01751	&0.45	&0.63	&820\\
N6334In-MM9	&17:20:54.863	&$-35$:45:07.13	&5.66	&7360	&51	&150	&345	&0.01751	&0.44	&0.67	&870\\
N6334In-MM10&17:20:54.925	&$-35$:45:06.41	&4.63	&6020	&54	&291	&345	&0.01751	&0.79	&0.71	&930\\
N6334In-MM11&17:20:54.973	&$-35$:45:07.00	&4.53	&5890	&55	&149	&345	&0.01751	&0.40	&0.57	&740\\
N6334In-MM12&17:20:55.056	&$-35$:45:07.09	&3.93	&5110	&57	&305	&345	&0.01751	&0.78	&0.87	&1130\\
N6334In-MM13&17:20:55.156	&$-35$:45:04.53	&1.25	&1630	&83	&420	&345	&0.01751	&0.70	&0.74	&960\\
N6334In-MM14\supe&17:20:55.245&$-35$:45:03.90&0.00	&0	    &98	&1470	&345	&0.01751	&2.06	&1.13	&1470\\
N6334In-MM15&17:20:55.621	&$-35$:45:07.46	&5.80	&7540	&50	&130	&345	&0.01751	&0.39	&0.69	&900\\
\hline
\end{tabular}
\begin{list}{}{}
\item[$^\mathrm{a}$] Separation from each fragment to the centre of the envelope (taken as the position of the brightest fragment, see Sec.~\ref{sec:Aina} and Appendix~\ref{app}). 
\item[$^\mathrm{b}$] Temperature at the position of the fragment obtained using the temperature power-law of the envelope given in Table~\ref{tab:cores1} at a distance $r_\mathrm{p}$ (this table), except for fragments taken as reference, for which the temperature was taken as $T_0$, the temperature at 1000 au (given in Table~\ref{tab:cores1}).
\item[$^\mathrm{c}$] Flux density, frequency of the interferometric observations where the fragments were identified, adopted dust opacity (from \citealt{Ossenkopf_Henning94}) and estimated mass of each fragment ($M_\mathrm{\delta r}$, using the temperature of each fragment, $T_\mathrm{env}$, given in this table).
\item[$^\mathrm{d}$] Effective radius estimated for each fragment taking the 3$\sigma$ contour level area, $A$, and using $\delta r/2=\sqrt{A/\pi}$.
\item[$^\mathrm{e}$] For each envelope listed in Table~\ref{tab:cores1}, this is the fragment with highest intensity, whose position (listed also in Table~\ref{tab:cores1}) is taken as the centre of the envelope. Fragment W3H2O-MM3 is an UCHII region and its millimetre emission corresponds to free-free emission.
\end{list}
\end{table*}

\begin{table*}
\contcaption{Position, mass and size of the 160 compact fragments embedded in the massive dense envelopes of \citet{Palau21}.}
\label{tab:continued}
\begin{tabular}{lccccccccccc}
\hline
&Fragment R.A. & Fragment Dec. &$r_\mathrm{p}$\supa &$r_\mathrm{p}$\supa &$T_{\rm env}$\supb &S$_\nu$\supc &Freq.\supc &Opacity\supc &$M_\mathrm{\delta r}$\supc &$\delta r/2$\supd &$\delta r/2$\supd \\         
Fragment   &(J2000) &(J2000) &(arcsec) &(au) &(K)   &(mJy)&(GHz) &($\si{\square\centi \meter\per\gram}$) &(\mo) &(arcsec) &(au)\\
\hline
G34-0-MM1	&18:53:17.471	&+01:25:30.30	&9.33	&14650	&25	&22	  &345	&0.01751	&0.22	&0.35	&550\\
G34-0-MM2	&18:53:17.832	&+01:25:28.65	&4.07	&6390	&34	&63	  &345	&0.01751	&0.45   &0.56	&890\\
G34-0-MM3\supe&18:53:18.005	&+01:25:25.51	&0.00	&0	    &63	&3830 &345	&0.01751	&12.8   &1.64	&2580\\
G34-0-MM4	&18:53:18.238	&+01:25:28.10	&4.35	&6830	&33	&46	  &345	&0.01751	&0.34	&0.50	&780\\
G34-0-MM5	&18:53:18.354	&+01:25:25.51	&5.23	&8220	&31	&32	  &345	&0.01751	&0.25	&0.43	&670\\
\hline
G34-1-MM1	&18:53:18.387	&+01:24:48.49	&8.44	&13260	&24	&13	&345	&0.01751	&0.14	&0.54	&850\\
G34-1-MM2	&18:53:18.505	&+01:24:43.46	&3.52	&5530	&33	&37	&345	&0.01751	&0.26	&0.57	&890\\
G34-1-MM3	&18:53:18.558	&+01:24:44.13	&3.40	&5340	&34	&156&345	&0.01751	&1.09	&0.96	&1510\\
G34-1-MM4	&18:53:18.572	&+01:24:40.57	&2.08	&3270	&41	&10	&345	&0.01751	&0.06	&0.44	&700\\
G34-1-MM5	&18:53:18.580	&+01:24:42.70	&2.17	&3400	&40	&147&345	&0.01751	&0.84	&0.87	&1370\\
G34-1-MM6	&18:53:18.678	&+01:24:42.28	&0.88	&1390	&56	&38	&345	&0.01751	&0.15	&0.52	&820\\
G34-1-MM7\supe&18:53:18.698	&+01:24:41.45	&0.00	&0	    &63	&138&345	&0.01751	&0.46	&0.75	&1180\\
G34-1-MM8	&18:53:18.728	&+01:24:35.16	&6.31	&9900	&27	&346&345	&0.01751	&3.26	&1.37	&2150\\
G34-1-MM9	&18:53:18.770	&+01:24:39.94	&1.86	&2910	&42	&94	&345	&0.01751	&0.50	&0.82	&1300\\
G34-1-MM10	&18:53:18.784	&+01:24:41.07	&1.34	&2110	&48	&41	&345	&0.01751	&0.19	&0.59	&930\\
\hline
G35-MM1	&18:58:12.619	&+01:40:38.40	&5.14	&11260	&40	&12	&345	&0.01751	&0.14	&0.46	&1020\\
G35-MM2	&18:58:12.807	&+01:40:40.08	&3.54	&7750	&45	&999&345	&0.01751	&9.66	&1.66	&3640\\
G35-MM3	&18:58:12.890	&+01:40:38.18	&1.30	&2840	&63	&144&345	&0.01751	&0.94	&0.72	&1570\\
G35-MM4\supe&18:58:12.954&+01:40:37.31	&0.00	&0	    &90	&838&345	&0.01751	&3.66	&1.16	&2540\\
G35-MM5	&18:58:13.034	&+01:40:35.92	&1.84	&4020	&56	&672&345	&0.01751	&5.00	&1.29	&2820\\
G35-MM6	&18:58:13.041	&+01:40:31.10	&6.35	&13900	&37	&47	&345	&0.01751	&0.58	&0.71	&1560\\
G35-MM7	&18:58:13.095	&+01:40:30.23	&7.39	&16180	&35	&19	&345	&0.01751	&0.25	&0.52	&1130\\
G35-MM8	&18:58:13.098	&+01:40:31.34	&6.35	&13900	&37	&65	&345	&0.01751	&0.80	&0.58	&1280\\
G35-MM9	&18:58:13.137	&+01:40:33.19	&4.95	&10840	&40	&82	&345	&0.01751	&0.91	&0.80	&1740\\
G35-MM10&18:58:13.185	&+01:40:31.45	&6.81	&14910	&36	&138&345	&0.01751	&1.75	&0.76	&1650\\
G35-MM11&18:58:13.210	&+01:40:30.84	&7.52	&16480	&35	&180&345	&0.01751	&2.38	&0.81	&1770\\
G35-MM12&18:58:13.291	&+01:40:30.86	&8.19	&17940	&34	&131&345	&0.01751	&1.80	&0.86	&1890\\
G35-MM13&18:58:13.333	&+01:40:31.95	&7.81	&17110	&34	&78	&345	&0.01751	&1.05	&0.73	&1590\\
G35-MM14&18:58:12.867	&+01:40:30.50	&6.93	&15180	&36	&21	&345	&0.01751	&0.27	&0.74	&1620\\
G35-MM15&18:58:12.967	&+01:40:30.61	&6.70	&14680	&36	&16	&345	&0.01751	&0.21	&0.66	&1460\\
\hline
I20126-MM1&20:14:26.030	&+41:13:32.56	&0.00	&0	    &86	&254&229	&0.00899	&2.80	&0.80	&1300\\
\hline
CygX-N3-MM1	&20:35:33.866	&+42:20:01.92	&7.89	&11040	&19	&6.7&229	&0.00899	&0.30	&0.70	&990\\
CygX-N3-MM2	&20:35:34.228	&+42:20:04.69	&3.07	&4300	&27	&12	&229	&0.00899	&0.34	&0.77	&1080\\
CygX-N3-MM3\supe&20:35:34.410&+42:20:07.00	&0.00	&0	    &45	&39	&229	&0.00899	&0.64	&1.11	&1560\\
CygX-N3-MM4	&20:35:34.465	&+42:20:09.83	&2.90	&4050	&28	&6.6&229	&0.00899	&0.19	&0.70	&970\\
CygX-N3-MM5	&20:35:34.554	&+42:20:00.20	&6.98	&9780	&20	&10	&229	&0.00899	&0.45	&0.82	&1150\\
CygX-N3-MM6	&20:35:34.634	&+42:20:08.72	&3.02	&4230	&27	&48	&229	&0.00899	&1.41	&1.12	&1570\\
\hline
W75N-MM1	&20:38:35.525	&+42:37:32.80	&10.76	&15060	&46	&40	&229	&0.00899	&0.64	&1.13	&1590\\
W75N-MM2	&20:38:35.787	&+42:37:26.40	&10.58	&14810	&46	&33	&229	&0.00899	&0.52	&0.99	&1390\\
W75N-MM3	&20:38:35.940	&+42:37:25.12	&10.40	&14550	&46	&23	&229	&0.00899	&0.37	&0.92	&1300\\
W75N-MM4	&20:38:35.986	&+42:37:26.53	&8.97	&12550	&49	&24	&229	&0.00899	&0.35	&0.89	&1250\\
W75N-MM5	&20:38:35.996	&+42:37:31.63	&5.84	&8170	&56	&644&229	&0.00899	&8.23	&2.33	&3260\\
W75N-MM6	&20:38:36.315	&+42:37:29.09	&4.85	&6780	&60	&61	&229	&0.00899	&0.73	&1.20	&1680\\
W75N-MM7	&20:38:36.394	&+42:37:22.04	&11.52	&16120	&45	&48	&229	&0.00899	&0.79	&1.40	&1960\\
W75N-MM8	&20:38:36.426	&+42:37:34.49	&1.26	&1770	&93	&819&229	&0.00899	&6.07	&2.26	&3160\\
W75N-MM9	&20:38:36.490	&+42:37:27.81	&5.69	&7970	&56	&53	&229	&0.00899	&0.67	&1.26	&1770\\
W75N-MM10\supe&20:38:36.497	&+42:37:33.50	&0.00	&0	    &112&478&229	&0.00899	&2.90	&1.49	&2090\\
W75N-MM11	&20:38:37.013	&+42:37:33.78	&5.70	&7990	&56	&38	&229	&0.00899	&0.48	&1.16	&1620\\
W75N-MM12	&20:38:37.038	&+42:37:35.17	&6.20	&8680	&55	&54	&229	&0.00899	&0.71	&1.09	&1530\\
W75N-MM13	&20:38:37.131	&+42:37:37.24	&7.94	&11110	&51	&236&229	&0.00899	&3.37	&1.79	&2510\\
W75N-MM14	&20:38:37.235	&+42:37:25.66	&11.31	&15830	&45	&117&229	&0.00899	&1.91	&1.83	&2570\\
\hline
\end{tabular}
\end{table*}

\begin{table*}
\contcaption{Position, mass and size of the 160 compact fragments embedded in the massive dense envelopes of \citet{Palau21}.}
\label{tab:continued2}
\begin{tabular}{lccccccccccc}
\hline
&Fragment R.A. & Fragment Dec. &$r_\mathrm{p}$\supa &$r_\mathrm{p}$\supa &$T_{\rm env}$\supb &S$_\nu$\supc &Freq.\supc &Opacity\supc &$M_\mathrm{\delta r}$\supc &$\delta r/2$\supd &$\delta r/2$\supd \\         
Fragment   &(J2000) &(J2000) &(arcsec) &(au) &(K)   &(mJy)&(GHz) &($\si{\square\centi \meter\per\gram}$) &(\mo) &(arcsec) &(au)\\
\hline
DR21OH-MM1	&20:39:00.399	&+42:22:45.62	&7.42	&10400	&31	&171&345	&0.01751	&1.05	&0.79	&1110\\
DR21OH-MM2	&20:39:00.461	&+42:22:44.66	&7.33	&10260	&32	&241&345	&0.01751	&1.47	&0.86	&1200\\
DR21OH-MM3	&20:39:00.469	&+42:22:46.87	&6.22	&8710	&33	&318&345	&0.01751	&1.80	&1.29	&1810\\
DR21OH-MM4	&20:39:00.596	&+42:22:43.10	&7.33	&10260	&32	&84	&345	&0.01751	&0.51	&0.65	&920\\
DR21OH-MM5	&20:39:00.612	&+42:22:42.77	&7.49	&10490	&31	&157&345	&0.01751	&0.97	&0.90	&1260\\
DR21OH-MM6	&20:39:00.645	&+42:22:47.23	&4.27	&5980	&38	&93	&345	&0.01751	&0.44	&0.82	&1150\\
DR21OH-MM7	&20:39:00.650	&+42:22:44.96	&5.53	&7740	&35	&174&345	&0.01751	&0.93	&0.80	&1130\\
DR21OH-MM8	&20:39:00.707	&+42:22:49.30	&3.26	&4570	&42	&56	&345	&0.01751	&0.24	&0.75	&1040\\
DR21OH-MM9	&20:39:00.818	&+42:22:50.71	&2.70	&3780	&45	&37	&345	&0.01751	&0.14	&0.58	&810\\
DR21OH-MM10	&20:39:00.910	&+42:22:41.93	&7.05	&9870	&32	&66	&345	&0.01751	&0.40	&0.82	&1140\\
DR21OH-MM11	&20:39:00.999	&+42:22:44.06	&4.85	&6790	&37	&48	&345	&0.01751	&0.24	&0.64	&900\\
DR21OH-MM12\supe&20:39:00.999&+42:22:48.91	&0.00	&0	    &73	&1090&345	&0.01751	&2.46	&1.54	&2150\\
DR21OH-MM13	&20:39:01.037	&+42:22:45.26	&3.67	&5140	&40	&35	&345	&0.01751	&0.16	&0.56	&790\\
DR21OH-MM14	&20:39:01.042	&+42:22:42.89	&6.04	&8450	&34	&65	&345	&0.01751	&0.36	&0.72	&1010\\
DR21OH-MM15	&20:39:01.080	&+42:22:44.51	&4.49	&6290	&38	&23	&345	&0.01751	&0.11	&0.44	&620\\
DR21OH-MM16	&20:39:01.137	&+42:22:49.12	&1.54	&2160	&55	&952&345	&0.01751	&2.94	&1.29	&1810\\
DR21OH-MM17	&20:39:01.207	&+42:22:51.31	&3.33	&4660	&42	&933&345	&0.01751	&4.00	&1.37	&1910\\
DR21OH-MM18	&20:39:01.329	&+42:22:51.37	&4.41	&6180	&38	&446&345	&0.01751	&2.16	&1.22	&1710\\
\hline
CygX-N48-MM1	&20:39:01.088	&+42:22:08.25	&4.47	&6260	&31	&89	&229	&0.00899	&2.22	&1.52	&2130\\
CygX-N48-MM2	&20:39:01.104	&+42:22:04.02	&2.72	&3810	&37	&11	&229	&0.00899	&0.22	&0.77	&1080\\
CygX-N48-MM3	&20:39:01.195	&+42:22:05.08	&1.64	&2300	&44	&12	&229	&0.00899	&0.20	&0.72	&1010\\
CygX-N48-MM4	&20:39:01.234	&+42:22:08.41	&3.84	&5370	&33	&39	&229	&0.00899	&0.92	&1.08	&1510\\
CygX-N48-MM5	&20:39:01.303	&+42:22:02.74	&2.06	&2890	&40	&30	&229	&0.00899	&0.55	&0.88	&1230\\
CygX-N48-MM6\supe&20:39:01.340	&+42:22:04.76	&0.00	&0	    &58	&95	&229	&0.00899	&1.17   &1.34	&1870\\
CygX-N48-MM7	&20:39:01.375	&+42:22:03.39	&1.42	&1990	&46	&12	&229	&0.00899	&0.20	&0.59	&820\\
CygX-N48-MM8	&20:39:01.677	&+42:22:02.25	&4.50	&6300	&31	&12	&229	&0.00899	&0.31	&0.81	&1140\\
CygX-N48-MM9	&20:39:01.739	&+42:22:01.84	&5.30	&7420	&29	&7.2&229	&0.00899	&0.19	&0.61	&850\\
CygX-N48-MM10	&20:39:01.914	&+42:21:59.98	&7.96	&11150	&26	&12	&229	&0.00899	&0.38	&0.84	&1180\\
CygX-N48-MM11	&20:39:01.926	&+42:22:02.85	&6.78	&9490	&27	&34	&229	&0.00899	&1.01	&1.25	&1740\\
CygX-N48-MM12	&20:39:02.096	&+42:22:04.32	&8.40	&11760	&25	&19	&229	&0.00899	&0.61	&0.92	&1290\\
\hline
CygX-N53-MM1	&20:39:02.697  &+42:25:42.96    &8.54	&11950	&18	&20	&229	&0.00899	&0.96	&0.81	&1140\\
CygX-N53-MM2	&20:39:02.800  &+42:25:55.16    &4.53	&6340	&23	&10	&229	&0.00899	&0.36	&0.61	&850\\
CygX-N53-MM3	&20:39:02.803  &+42:25:42.87    &8.30	&11620	&19	&11	&229	&0.00899	&0.52	&0.60	&840\\
CygX-N53-MM4	&20:39:02.844  &+42:25:40.70    &10.37	&14520	&17	&16	&229	&0.00899	&0.84	&0.78	&1100\\
CygX-N53-MM5	&20:39:02.927  &+42:25:45.13    &5.87	&8220	&21	&16	&229	&0.00899	&0.63	&0.77	&1080\\
CygX-N53-MM6\supe&20:39:02.959 &+42:25:50.99    &0.00	&0	    &45	&169&229	&0.00899	&2.76	&1.35	&1890\\
CygX-N53-MM7	&20:39:03.010  &+42:25:52.51    &1.62	&2270	&34	&33	&229	&0.00899	&0.76	&1.04	&1460\\
CygX-N53-MM8	&20:39:03.091  &+42:25:49.55    &2.05	&2870	&31	&20	&229	&0.00899	&0.51	&0.77	&1080\\
CygX-N53-MM9	&20:39:03.220  &+42:25:51.21    &2.90	&4060	&27	&115&229	&0.00899	&3.38	&1.29	&1810\\
\hline
CygX-N63-MM1\supe&20:40:05.378	&+41:32:13.04	&0.00	&0	    &45	&579&229	&0.00899	&9.44	&2.09	&2930\\
CygX-N63-MM2	&20:40:05.703	&+41:32:11.72	&3.88	&5440	&25	&17	&229	&0.00899	&0.55	&0.70	&970\\
\hline
N7538S-MM1	&23:13:43.893	&+61:26:50.19	&7.88	&20880	&32	&11	&218	&0.00798	&1.18	&0.36	&950\\
N7538S-MM2	&23:13:44.122	&+61:26:49.13	&6.26	&16580	&35	&3.8&218	&0.00798	&0.37	&0.29	&770\\
N7538S-MM3	&23:13:44.249	&+61:26:47.41	&5.81	&15410	&36	&2.7&218	&0.00798	&0.25	&0.25	&660\\
N7538S-MM4	&23:13:44.427	&+61:26:47.43	&4.66	&12360	&39	&7.1&218	&0.00798	&0.61	&0.30	&780\\
N7538S-MM5	&23:13:44.512	&+61:26:47.73	&3.99	&10560	&41	&54	&218	&0.00798	&4.36	&0.60	&1590\\
N7538S-MM6	&23:13:44.568	&+61:26:42.94	&7.47	&19790	&33	&9.8&218	&0.00798	&1.02	&0.43	&1140\\
N7538S-MM7	&23:13:44.743	&+61:26:48.62	&2.11	&5590	&51	&63	&218	&0.00798	&3.96	&0.79	&2090\\
N7538S-MM8\supe	&23:13:44.989   &+61:26:49.77	&0.00	&0	    &93	&95	&218	&0.00798	&3.12	&0.61	&1610\\
N7538S-MM9	&23:13:45.070	&+61:26:50.02	&0.63	&1680	&78	&35	&218	&0.00798	&1.39	&0.44	&1170\\
N7538S-MM10	&23:13:45.177	&+61:26:45.20	&4.76	&12630	&38	&5.3&218	&0.00798	&0.46	&0.33	&890\\
N7538S-MM11	&23:13:45.191	&+61:26:50.26	&1.53	&4060	&57	&4.0&218	&0.00798	&0.22	&0.30	&800\\
N7538S-MM12	&23:13:45.257	&+61:26:50.68	&2.13	&5640	&51	&3.6&218	&0.00798	&0.23	&0.28	&750\\
\hline
\end{tabular}
\end{table*}

\begin{table*}
\caption{Parameters of the 160 compact fragments studied in this work.}
\label{tab:fragments2}
\begin{tabular}{lcccccccccccc}
\hline
&$n_\mathrm{fluct}$\supa &$n_\mathrm{env}$\supb &  &$c_\mathrm{s,env}$\supb  & & $M_{\rm Jog}$\supd & $M_{\rm env}(r_\mathrm{p})$\supe  \\
Fragment &(cm$^{-3}$) &(cm$^{-3}$)   &$\eta$\supc &($\si{\kilo\meter\per\second}$)  &$M_\mathrm{\delta r}/M_{\rm Jog}$\supd    &(\mo) &(\mo) &${\cal{M}}_{\rm ff-turb}$\supf  \\
\hline
W3IRS5-MM1	&$8.58\times10^{5}$	&$4.70\times10^{5}$	&1.8	&0.67	&0.004	&18.46	&4.9&0.7\\
W3IRS5-MM2	&$8.20\times10^{5}$	&$2.90\times10^{6}$	&0.3	&0.86	&$-$	    &$-$	    &0.7&0.4\\
W3IRS5-MM3\supg	&$8.80\times10^{5}$	&$-$	                &$-$	    &0.95	&$-$	    &$-$    	&$-$	&$-$\\
W3IRS5-MM4	&$9.99\times10^{5}$	&$2.29\times10^{5}$	&4.4	&0.61	&0.01	&10.84	&10	&0.9\\
\hline
W3H2O-MM1	&$2.15\times10^{8}$	&$2.91\times10^{5}$	&739&0.45	&1.56	&0.26	&48	&2.3\\
W3H2O-MM2	&$2.34\times10^{8}$	&$2.69\times10^{5}$	&870&0.44	&1.56	&0.25	&51	&2.3\\
W3H2O-MM3\supg	&$-$	&$-$	&$-$	&0.73	&$-$	&$-$	&$-$	&$-$\\
W3H2O-MM4	&$1.17\times10^{8}$	&$3.15\times10^{6}$	&37	&0.57	&0.34	&0.76	&12	&1.7\\
W3H2O-MM5	&$9.56\times10^{7}$	&$1.10\times10^{6}$	&87	&0.51	&2.62	&0.60	&22	&1.9\\
W3H2O-MM6	&$1.53\times10^{8}$	&$7.92\times10^{5}$	&193&0.49	&0.64	&0.43	&27	&2.0\\
W3H2O-MM7	&$1.84\times10^{8}$	&$4.86\times10^{5}$	&380&0.47	&28.76	&0.33	&36	&2.2\\
W3H2O-MM8	&$1.50\times10^{8}$	&$3.36\times10^{5}$	&445&0.45	&33.21	&0.33	&44	&2.3\\
\hline
G192-MM1    &$1.06\times10^{7}$ &$-$  &$-$   &0.48   &$-$ &$-$       &$-$   &$-$  \\      
\hline
N6334V-MM1	&$6.98\times10^{6}$	&$1.82\times10^{6}$	&3.8&0.45	&0.07	&2.16	&14	&2.0\\
N6334V-MM2\supg	&$9.53\times10^{6}$	&$-$ &$-$	&0.58	&$-$	    &$-$	    &$-$	&$-$\\
N6334V-MM3	&$1.08\times10^{7}$	&$4.98\times10^{6}$	&2.2&0.49	&0.39	&3.15	&7.5&1.8\\
N6334V-MM4	&$1.02\times10^{7}$	&$7.77\times10^{5}$	&13	&0.42	&0.11	&1.12	&22	&2.2\\
N6334V-MM5	&$8.60\times10^{6}$	&$8.30\times10^{5}$	&10	&0.42	&0.52	&1.27	&22	&2.2\\
\hline
N6334A-MM1	&$2.90\times10^{7}$	&$2.37\times10^{6}$	&12	&0.44	&0.67	&0.69	&3.6&1.3\\
N6334A-MM2	&$1.23\times10^{7}$	&$7.77\times10^{6}$	&1.6&0.50	&1.05	&2.02	&1.0&0.9\\
N6334A-MM3	&$2.75\times10^{7}$	&$2.90\times10^{6}$	&9.5&0.45	&0.47	&0.77	&2.9&1.2\\
N6334A-MM4\supg	&$1.26\times10^{7}$	&$-$ &$-$	&0.52	&$-$	    &$-$    	&$-$	&$-$\\
N6334A-MM5	&$8.83\times10^{6}$	&$1.01\times10^{6}$	&8.7&0.40	&0.10	&0.95	&9.3&1.7\\
N6334A-MM6	&$9.87\times10^{6}$	&$1.82\times10^{6}$	&5.4&0.42	&1.11	&1.13	&4.8&1.4\\
N6334A-MM7	&$1.10\times10^{7}$	&$2.34\times10^{5}$	&47	&0.34	&0.53	&0.50	&48	&2.8\\
N6334A-MM8	&$1.46\times10^{7}$	&$4.39\times10^{5}$	&33	&0.36	&1.67	&0.54	&24	&2.3\\
N6334A-MM9	&$2.22\times10^{7}$	&$2.68\times10^{5}$	&83	&0.34	&3.16	&0.37	&41	&2.6\\
N6334A-MM10	&$2.06\times10^{7}$	&$2.70\times10^{5}$	&76	&0.34	&9.78	&0.38	&41	&2.6\\
N6334A-MM11	&$2.42\times10^{7}$	&$2.24\times10^{5}$	&108&0.33	&6.20	&0.33	&50	&2.8\\
N6334A-MM12	&$1.52\times10^{7}$	&$1.93\times10^{5}$	&79	&0.33	&1.85	&0.40	&59	&2.9\\
N6334A-MM13	&$1.32\times10^{7}$	&$1.65\times10^{5}$	&80	&0.32	&1.67	&0.40	&70	&3.1\\
N6334A-MM14	&$2.20\times10^{7}$	&$1.26\times10^{5}$	&175&0.31	&1.80	&0.28	&95	&3.4\\
N6334A-MM15	&$1.91\times10^{7}$	&$1.14\times10^{5}$	&167&0.31	&2.43	&0.30	&106&3.5\\
N6334A-MM16	&$2.85\times10^{7}$	&$1.11\times10^{5}$	&256&0.31	&14.34	&0.24	&109&3.5\\
\hline
N6334I-MM1	&$3.21\times10^{7}$	&$3.12\times10^{5}$	&103&0.42	&0.26	&0.56	&63	&2.8\\
N6334I-MM2	&$2.21\times10^{7}$	&$8.24\times10^{5}$	&27	&0.45	&0.23	&0.88	&39	&2.6\\
N6334I-MM3	&$3.97\times10^{7}$	&$1.28\times10^{6}$	&31	&0.46	&0.70	&0.72	&32	&2.5\\
N6334I-MM4	&$3.05\times10^{7}$	&$6.78\times10^{5}$	&45	&0.44	&0.90	&0.70	&43	&2.6\\
N6334I-MM5	&$2.01\times10^{7}$	&$3.69\times10^{5}$	&54	&0.42	&0.22	&0.75	&58	&2.7\\
N6334I-MM6	&$3.58\times10^{7}$	&$2.96\times10^{6}$	&12	&0.50	&5.98	&1.01	&21	&2.4\\
N6334I-MM7\supg	&$3.35\times10^{7}$	&$-$ &$-$	&0.62	&$-$	&$-$	&$-$	&$-$\\
\hline
N6334In-MM1	&$3.20\times10^{7}$	&$2.52\times10^{5}$	&127&0.35	&0.97	&0.34	&137&3.8\\
N6334In-MM2	&$4.08\times10^{7}$	&$2.50\times10^{5}$	&163&0.35	&5.86	&0.30	&138&3.8\\
N6334In-MM3	&$3.13\times10^{7}$	&$2.96\times10^{5}$	&106&0.36	&4.21	&0.36	&115&3.6\\
N6334In-MM4	&$5.71\times10^{7}$	&$2.83\times10^{5}$	&201&0.36	&7.04	&0.26	&121&3.7\\
N6334In-MM5	&$2.15\times10^{7}$	&$7.12\times10^{5}$	&30	&0.40	&0.90	&0.59	&46	&2.8\\
N6334In-MM6	&$3.15\times10^{7}$	&$3.65\times10^{5}$	&86	&0.37	&1.41	&0.38	&92	&3.4\\

\hline
\end{tabular}
\begin{list}{}{}
\item[$^\mathrm{a}$] Number density of the fluctuation (fragment) estimated as $n_\mathrm{fluct} = \frac{3\,M_\mathrm{\delta r}}{\mu\,m_\mathrm{H}\,4\pi (\delta r/2)^{3}}$, where $\mu$ is the mean molecular weight, taken as 2.36, and $m_\mathrm{H}$ is the mass of the Hydrogen atom. $M_\mathrm{\delta r}$ and $\delta r/2$ are taken from Table~\ref{tab:fragments1}. 
\item[$^\mathrm{c}$] Properties of the envelope at the position of each fragment. $c_\mathrm{s,env}$ is the sound speed at the position of the fragment, estimated as $c_\mathrm{s,env}=\sqrt{k_\mathrm{B}T_\mathrm{env}/(\mu\,m_\mathrm{H})}$, where $k_\mathrm{B}$ is the Boltzmann constant, and $T_\mathrm{env}$ is the temperature of the envelope at the position of the fragment (Table~\ref{tab:fragments1}). $n_\mathrm{env}$ is the number density of the envelope at the position of the fragment, calculated using the density power law (given in Table~\ref{tab:cores1}) for a distance $r_\mathrm{p}$ (given in Table~\ref{tab:fragments1}): $n_{\rm env}=\rho_\mathrm{1000} (r_\mathrm{p}/r_\mathrm{1000})^{-\alpha}/(\mu\,m_\mathrm{H})$.
\item[$^\mathrm{c}$] Amplitude of the fluctuation or density contrast, calculated as $\eta= n_{\rm fluct}/ n_{\rm env}$, following eq.~\ref{eq:den_pert_potencias}.
\item[$^\mathrm{d}$] Ratio of the mass of the fluctuation (fragment) over the Jog mass calculated following eq.~\ref{eq:criterio:colapso}. All the parameters required by this equation are listed in Tables~\ref{tab:fragments1} and \ref{tab:fragments2}, with $n_0=n_\mathrm{env}$ (the density of the envelope at the position of the fragment). $M_\mathrm{Jog}$ is calculated as $M_\mathrm{Jog}=M_\mathrm{\delta r}/(M_\mathrm{\delta r}/M_{\rm Jog})$.
\item[$^\mathrm{e}$] Mass of the envelope enclosed within a radius $r_\mathrm{p}$ (the position of each fragment), calculated following $M_\mathrm{env}(r_\mathrm{p})=4\pi\rho_\mathrm{1000}\,r_\mathrm{1000}^\alpha\,\frac{r_\mathrm{p}^{3-\alpha}}{3-\alpha}$ (see Table~3 of \citealt{Palau21}), where $\rho_\mathrm{1000}$ is given in Table~\ref{tab:cores1} and $r_\mathrm{1000}=1000$~au. 
\item[$^\mathrm{f}$] Mach number of the turbulence generated by the gravitational collapse of the envelope, calculated as $\Machffturb = \frac{\epsilon_{\rm ff-turb}}{c_\mathrm{s,env}} \sqrt{\frac{2 G \Menv(r_{\mathrm{p}})}{r_\mathrm{p}}}$ (equation \ref{eq:Machffturb}), where $c_\mathrm{s,env}$, and $\Menv(r_\mathrm{p})$ are given in this table, and $\epsilon_{\rm ff-turb}$ is adopted to be 0.4 (Sec.~\ref{sec:formationfragments}).
\item[$^\mathrm{g}$] Fragment with highest intensity, whose position (listed in Table~\ref{tab:cores1}) is taken as the centre of the envelope. Fragment W3H2O-MM3 is an UCHII region.
\end{list}
\end{table*}

\begin{table*}
\contcaption{Parameters of the 160 compact fragments studied in this work.}
\label{tab:continued3}
\begin{tabular}{lcccccccccccc}
\hline
&$n_\mathrm{fluct}$\supa &$n_\mathrm{env}$\supb &  &$c_\mathrm{s,env}$\supb  & & $M_{\rm Jog}$\supd & $M_{\rm env}(r_\mathrm{p})$\supe  \\
Fragment &(cm$^{-3}$) &(cm$^{-3}$)   &$\eta$\supc &($\si{\kilo\meter\per\second}$)  &$M_\mathrm{\delta r}/M_{\rm Jog}$\supd    &(\mo) &(\mo) &${\cal{M}}_{\rm ff-turb}$\supf  \\
\hline
N6334In-MM7	&$1.74\times10^{7}$	&$9.17\times10^{5}$	&19	&0.41	&1.15	&0.72	&35	&2.6\\
N6334In-MM8	&$2.94\times10^{7}$	&$4.05\times10^{5}$	&73	&0.37	&1.09	&0.41	&83	&3.3\\
N6334In-MM9	&$2.40\times10^{7}$	&$1.21\times10^{6}$	&20	&0.42	&0.66	&0.67	&26	&2.4\\
N6334In-MM10&$3.56\times10^{7}$ &$1.62\times10^{6}$	&22	&0.44	&1.30	&0.61	&19	&2.2\\
N6334In-MM11&$3.50\times10^{7}$	&$1.67\times10^{6}$	&21	&0.44	&0.65	&0.62	&19	&2.2\\
N6334In-MM12&$1.95\times10^{7}$	&$2.06\times10^{6}$	&9.5&0.45	&0.85	&0.92	&15	&2.0\\
N6334In-MM13&$2.88\times10^{7}$	&$1.09\times10^{7}$	&2.6&0.54	&0.46	&1.52	&2.6&1.2\\
N6334In-MM14\supg&$2.31\times10^{7}$	&$-$ &$-$	&0.59	&$-$	    &$-$	&$-$	&$-$\\
N6334In-MM15&$1.90\times10^{7}$	&$1.17\times10^{6}$	&16	&0.42	&0.51	&0.75	&27	&2.4\\
\hline
G34-0-MM1	&$4.77\times10^{7}$	&$1.17\times10^{5}$	&406&0.30	&1.36	&0.17	&42	&3.0\\
G34-0-MM2	&$2.32\times10^{7}$	&$7.65\times10^{5}$	&30	&0.34	&1.14	&0.39	&23	&2.9\\
G34-0-MM3\supg	&$2.68\times10^{7}$	&$-$	                &$-$	&0.47	&$-$	&$-$	&$-$	&$-$\\
G34-0-MM4	&$2.57\times10^{7}$	&$6.59\times10^{5}$	&39	&0.34	&0.95	&0.35	&24	&2.9\\
G34-0-MM5	&$3.05\times10^{7}$	&$4.34\times10^{5}$	&70	&0.33	&0.88	&0.29	&27	&3.0\\
\hline
G34-1-MM1	&$7.93\times10^{6}$	&$2.20\times10^{5}$	&36	&0.29	&0.36	&0.39	&34	&3.0\\
G34-1-MM2	&$1.32\times10^{7}$	&$1.02\times10^{6}$	&13	&0.34	&0.52	&0.51	&12	&2.3\\
G34-1-MM3	&$1.15\times10^{7}$	&$1.09\times10^{6}$	&11	&0.34	&1.93	&0.57	&11	&2.2\\
G34-1-MM4	&$6.11\times10^{6}$	&$2.58\times10^{6}$	&2.4&0.38	&0.04	&1.46	&6.1&1.9\\
G34-1-MM5	&$1.16\times10^{7}$	&$2.41\times10^{6}$	&4.8&0.37	&1.03	&0.81	&6.4&2.0\\
G34-1-MM6	&$9.63\times10^{6}$	&$1.17\times10^{7}$	&0.8&0.44	&0.02	&9.27	&2.1&1.5\\
G34-1-MM7\supg	&$1.00\times10^{7}$	&$-$	                &$-$	&0.47	&$-$	    &$-$	    &$-$	 &$-$\\
G34-1-MM8	&$1.18\times10^{7}$	&$3.67\times10^{5}$	&32	&0.31	&8.69	&0.38	&24	&2.7\\
G34-1-MM9	&$8.26\times10^{6}$	&$3.16\times10^{6}$	&2.6&0.39	&0.39	&1.28	&5.3&1.9\\
G34-1-MM10	&$8.37\times10^{6}$	&$5.58\times10^{6}$	&1.5&0.41	&0.08	&2.24	&3.5&1.7\\
\hline
G35-MM1	&$4.68\times10^{6}$	&$2.74\times10^{5}$	&17	&0.37	&0.13	&1.06	&22	&2.0\\
G35-MM2	&$7.21\times10^{6}$&$4.85\times10^{5}$	&15	&0.40	&9.31	&1.04	&13	&1.7\\
G35-MM3	&$8.64\times10^{6}$&$2.26\times10^{6}$	&3.8	&0.47	&0.53	&1.77	&2.9	&1.2\\
G35-MM4\supg	&$8.07\times10^{6}$	&$-$	&$-$	&0.56	&$-$	&$-$	&$-$	&$-$\\
G35-MM5	&$8.02\times10^{6}$	&$1.33\times10^{6}$	&6	&0.44	&3.44	&1.45	&4.9	&1.3\\
G35-MM6	&$5.50\times10^{6}$	&$1.99\times10^{5}$	&28	&0.36	&0.67	&0.87	&30	&2.2\\
G35-MM7	&$6.17\times10^{6}$	&$1.57\times10^{5}$	&39	&0.35	&0.33	&0.75	&38	&2.3\\
G35-MM8	&$1.37\times10^{7}$	&$1.99\times10^{5}$	&69	&0.36	&1.48	&0.54	&30	&2.2\\
G35-MM9	&$6.17\times10^{6}$	&$2.91\times10^{5}$	&21	&0.37	&0.97	&0.93	&21	&2.0\\
G35-MM10	&$1.39\times10^{7}$	&$1.79\times10^{5}$	&78	&0.35	&3.37	&0.52	&34	&2.3\\
G35-MM11	&$1.54\times10^{7}$	&$1.53\times10^{5}$	&101	&0.35	&5.10	&0.47	&39	&2.4\\
G35-MM12	&$9.50\times10^{6}$	&$1.34\times10^{5}$	&71	&0.34	&3.15	&0.57	&44	&2.4\\
G35-MM13	&$9.34\times10^{6}$	&$1.45\times10^{5}$	&65	&0.35	&1.77	&0.59	&41	&2.4\\
G35-MM14	&$2.25\times10^{6}$	&$1.74\times10^{5}$	&13	&0.35	&0.20	&1.32	&35	&2.3\\
G35-MM15	&$2.42\times10^{6}$	&$1.83\times10^{5}$	&13	&0.36	&0.16	&1.30	&33	&2.2\\
\hline
I20126&  $4.52\times10^{7}$     &$-$	               &$-$	&0.55	&$-$	&$-$	&$-$	&$-$\\
\hline
CygX-N3-MM1	&$1.12\times10^{7}$	&$2.28\times10^{5}$	&49	&0.26	&1.30	&0.23	&18	&2.6\\
CygX-N3-MM2	&$9.69\times10^{6}$	&$1.01\times10^{6}$	&10	&0.31	&0.80	&0.43	&4.7	&1.8\\
CygX-N3-MM3\supg	&$6.07\times10^{6}$	&$-$	&$-$	&0.40	&$-$	&$-$	&$-$	&$-$\\
CygX-N3-MM4	&$7.39\times10^{6}$	&$1.11\times10^{6}$	&6.7	&0.31	&0.36	&0.52	&4.3	&1.8\\
CygX-N3-MM5	&$1.05\times10^{7}$	&$2.76\times10^{5}$	&38	&0.27	&1.75	&0.25	&15	&2.5\\
CygX-N3-MM6	&$1.30\times10^{7}$	&$1.04\times10^{6}$	&13	&0.31	&3.84	&0.37	&4.6	&1.8\\
\hline
W75N-MM1	&$5.74\times10^{6}$	&$1.61\times10^{5}$	&36	&0.40	&0.53	&1.20	&45	&2.3\\
W75N-MM2	&$7.08\times10^{6}$	&$1.66\times10^{5}$	&43	&0.40	&0.48	&1.09	&45	&2.3\\
W75N-MM3	&$6.11\times10^{6}$	&$1.72\times10^{5}$	&36	&0.40	&0.31	&1.19	&44	&2.3\\
W75N-MM4	&$6.55\times10^{6}$ &$2.31\times10^{5}$	&28	&0.41	&0.28	&1.25	&38	&2.2\\
W75N-MM5	&$8.52\times10^{6}$	&$5.42\times10^{5}$	&16	&0.44	&5.83	&1.41	&24	&2.1\\
W75N-MM6	&$5.56\times10^{6}$	&$7.85\times10^{5}$	&7.1&0.46	&0.34	&2.17	&20	&2.0\\
W75N-MM7	&$3.76\times10^{6}$	&$1.40\times10^{5}$	&27	&0.40	&0.54	&1.46	&49	&2.3\\
W75N-MM8	&$6.89\times10^{6}$	&$1.14\times10^{7}$	&0.6&0.57	&$-$	&$-$	&5.2	&1.6\\
W75N-MM9	&$4.37\times10^{6}$	&$5.70\times10^{5}$	&7.7&0.44	&0.30	&2.23	&24	&2.1\\
W75N-MM10\supg	&$1.14\times10^{7}$	&$-$	&$-$	&0.63	&$-$	&$-$	&$-$	&$-$\\
W75N-MM11	&$4.08\times10^{6}$	&$5.68\times10^{5}$	&7.2&0.44	&0.21	&2.33	&24	&2.1\\
W75N-MM12	&$7.11\times10^{6}$	&$4.80\times10^{5}$	&15	&0.44	&0.47	&1.51	&26	&2.1\\
W75N-MM13	&$7.65\times10^{6}$	&$2.94\times10^{5}$	&26	&0.42	&2.74	&1.23	&33	&2.2\\
W75N-MM14	&$4.06\times10^{6}$	&$1.45\times10^{5}$	&28	&0.40	&1.35	&1.41	&48	&2.3\\
\hline
\end{tabular}
\end{table*}

\begin{table*}
\contcaption{Parameters of the 160 compact fragments studied in this work.}
\label{tab:continued4}
\begin{tabular}{lcccccccccccc}
\hline
&$n_\mathrm{fluct}$\supa &$n_\mathrm{env}$\supb &  &$c_\mathrm{s,env}$\supb  & & $M_{\rm Jog}$\supd & $M_{\rm env}(r_\mathrm{p})$\supe  \\
Fragment &(cm$^{-3}$) &(cm$^{-3}$)   &$\eta$\supc &($\si{\kilo\meter\per\second}$)  &$M_\mathrm{\delta r}/M_{\rm Jog}$\supd    &(\mo) &(\mo) &${\cal{M}}_{\rm ff-turb}$\supf  \\
\hline
DR21OH-MM1	&$2.76\times10^{7}$	&$7.61\times10^{5}$	&36	&0.33	&3.36	&0.31	&70	&4.2\\
DR21OH-MM2	&$3.05\times10^{7}$	&$7.82\times10^{5}$	&39	&0.33	&4.93	&0.30	&69	&4.2\\
DR21OH-MM3	&$1.09\times10^{7}$	&$1.08\times10^{6}$	&10	&0.34	&2.96	&0.61	&58	&4.0\\
DR21OH-MM4	&$2.40\times10^{7}$	&$7.81\times10^{5}$	&31	&0.33	&1.51	&0.34	&69	&4.2\\
DR21OH-MM5	&$1.73\times10^{7}$	&$7.48\times10^{5}$	&23	&0.33	&2.41	&0.40	&71	&4.2\\
DR21OH-MM6	&$1.04\times10^{7}$	&$2.27\times10^{6}$	&4.6&0.37	&0.47	&0.94	&40	&3.8\\
DR21OH-MM7	&$2.35\times10^{7}$	&$1.36\times10^{6}$	&17	&0.35	&2.25	&0.41	&52	&3.9\\
DR21OH-MM8	&$7.50\times10^{6}$	&$3.88\times10^{6}$	&1.9&0.38	&0.09	&2.51	&30	&3.6\\
DR21OH-MM9	&$9.81\times10^{6}$	&$5.66\times10^{6}$	&1.7&0.40	&0.05	&2.92	&25	&3.4\\
DR21OH-MM10	&$9.49\times10^{6}$	&$8.44\times10^{5}$	&11	&0.33	&0.66	&0.60	&66	&4.1\\
DR21OH-MM11	&$1.18\times10^{7}$	&$1.77\times10^{6}$	&6.7&0.36	&0.33	&0.73	&45	&3.8\\
DR21OH-MM12\supg	&$8.85\times10^{6}$	&$-$	&$-$	&0.51	&$-$	&$-$	&$-$	&$-$\\
DR21OH-MM13	&$1.14\times10^{7}$	&$3.07\times10^{6}$	&3.7&0.38	&0.15	&1.06	&34	&3.7\\
DR21OH-MM14	&$1.26\times10^{7}$	&$1.15\times10^{6}$	&11	&0.34	&0.64	&0.57	&57	&4.0\\
DR21OH-MM15	&$1.65\times10^{7}$	&$2.06\times10^{6}$	&8.0&0.36	&0.18	&0.61	&42	&3.8\\
DR21OH-MM16	&$1.78\times10^{7}$	&$1.70\times10^{7}$	&1.0&0.44	&0.08	&36.2	&14	&3.1\\
DR21OH-MM17	&$2.05\times10^{7}$	&$3.73\times10^{6}$	&5.5&0.38	&5.59	&0.72	&31	&3.6\\
DR21OH-MM18	&$1.56\times10^{7}$	&$2.13\times10^{6}$	&7.3&0.36	&3.31	&0.65	&41	&3.8\\
\hline
CygX-N48-MM1	&$8.26\times10^{6}$	&$1.10\times10^{6}$	&7.5	&0.33	&3.68	&0.60	&17	&2.7\\
CygX-N48-MM2	&$6.14\times10^{6}$	&$2.57\times10^{6}$	&2.4	&0.36	&0.18	&1.19	&9.2&2.3\\
CygX-N48-MM3	&$6.88\times10^{6}$	&$6.11\times10^{6}$	&1.1	&0.39	&0.07	&2.69	&4.8&2.0\\
CygX-N48-MM4	&$9.63\times10^{6}$	&$1.43\times10^{6}$	&6.7	&0.34	&1.49	&0.61	&14	&2.6\\
CygX-N48-MM5	&$1.05\times10^{7}$	&$4.13\times10^{6}$	&2.5	&0.38	&0.54	&1.02	&6.4&2.1\\
CygX-N48-MM6\supg	&$6.43\times10^{6}$	&$-$	&$-$	&0.45	&$-$	&$-$	&$-$	&$-$\\
CygX-N48-MM7	&$1.28\times10^{7}$	&$7.78\times10^{6}$	&1.6	&0.40	&0.14	&1.43	&4.0&1.9\\
CygX-N48-MM8	&$7.60\times10^{6}$	&$1.09\times10^{6}$	&7.0   &0.33	&0.50	&0.63	&18	&2.7\\
CygX-N48-MM9	&$1.12\times10^{7}$	&$8.22\times10^{5}$	&14	   &0.32	&0.43	&0.45	&22	&2.8\\
CygX-N48-MM10	&$8.34\times10^{6}$	&$4.10\times10^{5}$	&20	  &0.30	&0.92	&0.42	&37	&3.2\\
CygX-N48-MM11	&$6.81\times10^{6}$	&$5.40\times10^{5}$	&13	  &0.31	&1.96	&0.51	&30	&3.1\\
CygX-N48-MM12	&$1.02\times10^{7}$	&$3.74\times10^{5}$	&27	  &0.30	&1.69	&0.36	&39	&3.3\\
\hline
CygX-N53-MM1	&$2.32\times10^{7}$	&$3.12\times10^{5}$	&74	&0.25	&6.45	&0.15	&36	&3.6\\
CygX-N53-MM2	&$2.11\times10^{7}$	&$9.53\times10^{5}$	&22	&0.28	&1.60	&0.23	&16	&3.0\\
CygX-N53-MM3	&$3.21\times10^{7}$	&$3.28\times10^{5}$	&98	&0.26	&4.09	&0.13	&35	&3.6\\
CygX-N53-MM4	&$2.31\times10^{7}$	&$2.22\times10^{5}$	&104&0.25	&6.34	&0.13	&46	&3.9\\
CygX-N53-MM5	&$1.79\times10^{7}$	&$6.03\times10^{5}$	&30	&0.27	&3.00	&0.21	&23	&3.3\\
CygX-N53-MM6\supg	&$1.47\times10^{7}$	&$-$	&$-$	&0.40	&$-$	&$-$	&$-$	&$-$\\
CygX-N53-MM7	&$8.79\times10^{6}$	&$5.80\times10^{6}$	&1.5&0.34	&0.59	&1.27	&4.6&2.2\\
CygX-N53-MM8	&$1.44\times10^{7}$	&$3.83\times10^{6}$	&3.7&0.33	&0.97	&0.53	&6.1&2.4\\
CygX-N53-MM9	&$2.05\times10^{7}$	&$2.09\times10^{6}$	&10	&0.31	&10.99	&0.31	&9.4&2.6\\
\hline
CygX-N63-MM1\supg	&$1.35\times10^{7}$	&$-$	&$-$	&0.40	&$-$	&$-$	&$-$	&$-$\\
CygX-N63-MM2	&$2.15\times10^{7}$	&$5.30\times10^{5}$	&41	&0.30	&2.17	&0.25	&7.3	&2.1\\
\hline
N7538S-MM1	&$4.92\times10^{7}$	&$1.77\times10^{5}$	&278	&0.34	&5.08	&0.23	&105&3.6\\
N7538S-MM2	&$2.94\times10^{7}$ &$2.63\times10^{5}$	&112	&0.35	&1.08	&0.34	&78	&3.3\\
N7538S-MM3	&$3.13\times10^{7}$	&$2.98\times10^{5}$	&105	&0.35	&0.74	&0.34	&71	&3.2\\
N7538S-MM4	&$4.59\times10^{7}$	&$4.36\times10^{5}$	&105	&0.37	&1.92	&0.32	&54	&3.0\\
N7538S-MM5	&$3.89\times10^{7}$	&$5.71\times10^{5}$	&68	     &0.38	&11.56	&0.38	&44	&2.9\\
N7538S-MM6	&$2.49\times10^{7}$	&$1.94\times10^{5}$	&128	&0.34	&3.02	&0.34	&98	&3.5\\
N7538S-MM7	&$1.55\times10^{7}$	&$1.71\times10^{6}$	&9.1	&0.42	&4.37	&0.91	&19	&2.4\\
N7538S-MM8\supg	&$2.69\times10^{7}$	&$-$	&$-$	&0.57	&$-$	&$-$	&$-$	&$-$\\
N7538S-MM9	&$3.13\times10^{7}$	&$1.35\times10^{7}$	&2.3	&0.52	&0.84	&1.65	&4.2	&1.6\\
N7538S-MM10	&$2.38\times10^{7}$	&$4.20\times10^{5}$	&57	&0.37	&1.04	&0.44	&55	&3.0\\
N7538S-MM11	&$1.57\times10^{7}$	&$2.96\times10^{6}$	&5.3	&0.45	&0.19	&1.14	&13	&2.1\\
N7538S-MM12	&$1.93\times10^{7}$	&$1.68\times10^{6}$&12	&0.42	&0.29	&0.79	&20	&2.4\\
\hline
\end{tabular}
\end{table*}


\bsp	
\label{lastpage}
\end{document}